\documentclass[preprintnumbers, floatfix,letterpaper,aps,prd,epsfig,nofootinbib, twocolumn]{revtex4-1}

\usepackage{graphicx}% Include figure files
\usepackage{dcolumn}% Align table columns on decimal point
\usepackage[colorlinks,linkcolor=blue,citecolor=blue,urlcolor=blue]{hyperref}
\usepackage{bm,graphicx,dcolumn,epstopdf,epsf, latexsym,mathbbol, amssymb,amsmath, color, slashed, mathrsfs,mathcomp,simplewick}
\usepackage[center]{subfigure}
\usepackage{multirow}
\usepackage{makecell}

 \newcommand{\bq}{\begin{equation}}
 \newcommand{\eq}{\end{equation}}
 \newcommand{\bqn}{\begin{eqnarray}}
 \newcommand{\eqn}{\end{eqnarray}}
 \newcommand{\nb}{\nonumber}

\begin{document}

\title{Constraints on parity and Lorentz violations from gravitational waves: a comparison between single-parameter and multi-parameter analysis}

\author{Wei-Hua Guo}
%\email{guowh@zjut.edu.cn}
\affiliation{
Institute for theoretical physics \& cosmology, Zhejiang University of Technology, Hangzhou, 310032, China}
\affiliation{United Center for Gravitational Wave Physics, Zhejiang University of Technology, Hangzhou, 310032, China}

\author{Yuan-Zhu Wang}
\email{Corresponding author: vamdrew@zjut.edu.cn}
\affiliation{
Institute for theoretical physics \& cosmology, Zhejiang University of Technology, Hangzhou, 310032, China}
\affiliation{United Center for Gravitational Wave Physics, Zhejiang University of Technology, Hangzhou, 310032, China}

\author{Tao Zhu}
\email{Corresponding author: zhut05@zjut.edu.cn}
\affiliation{
Institute for theoretical physics \& cosmology, Zhejiang University of Technology, Hangzhou, 310032, China}
\affiliation{United Center for Gravitational Wave Physics, Zhejiang University of Technology, Hangzhou, 310032, China}

\author{Hai-Tian Wang}
\affiliation{
Institute for theoretical physics \& cosmology, Zhejiang University of Technology, Hangzhou, 310032, China}
\affiliation{United Center for Gravitational Wave Physics, Zhejiang University of Technology, Hangzhou, 310032, China}

\date{\today}

\begin{abstract}

The growing catalog of gravitational wave (GW) detections by the LIGO-Virgo-KAGRA Collaboration enables increasingly stringent tests of general relativity, particularly regarding possible {violations of parity and Lorentz symmetry}. Parity and Lorentz violations in gravity can modify both the damping rate and dispersion relation of GWs, leading to birefringence, frequency-dependent damping, and dispersion effects in the propagation of GWs. These effects result in amplitude and phase corrections of the waveforms of GWs produced by the coalescence of compact binaries, which enable us to constrain parity- and Lorentz-violating effects by analyzing GW signals detected by LIGO-Virgo-KAGRA detectors with the distorted waveforms. While most current analyses employ single-parameter methods—varying one deformation parameter at a time—modified gravity theories often predict multiple, coexisting deviations. In this work, we construct {several} specific multi-parameter GW waveform models incorporating parity- and Lorentz-violating effects and perform full Bayesian parameter estimation to compare multi-parameter and single-parameter constraints. We find that including multiple deformation parameters yields constraints on individual parameters that are generally comparable to those from single-parameter analyses, despite one specific model showing a degeneracy between the deformation parameters. Our results support the robustness of single-parameter tests for parity and Lorentz symmetry of gravity in current and future GW observations.

\end{abstract}

\maketitle

\section{Introduction}\label{sec:intro}
\renewcommand{\theequation}{1.\arabic{equation}} \setcounter{equation}{0}

In recent years, the availability of detection data on gravitational waves (GWs) from the coalescence of compact binary systems by the LIGO-Virgo-KAGRA Collaboration (LVK) has increased significantly \cite{LIGOScientific:2016emj, LIGOScientific:2016vlm, LIGOScientific:2016vbw, LIGOScientific:2016aoc, LIGOScientific:2017vwq, LIGOScientific:2018mvr, LIGOScientific:2020ibl, KAGRA:2021vkt}, making the study of GW physics a topic of growing interest in the scientific community. These GWs carry crucial information about the local spacetime properties of compact binaries in the strong gravitational field and highly dynamical spacetime region, enabling probing the nature of gravitational interaction in these extreme environments. Remarkably, most of the current GW observations have been shown to be in excellent agreement with the theoretical predictions of Einstein's theory of general relativity (GR) \cite{LIGOScientific:2019fpa, LIGOScientific:2020tif, LIGOScientific:2021sio}. The future ground- and space-based GW detectors, including the Einstein telescope \cite{Branchesi:2023mws}, Cosmic Explorer \cite{Evans:2021gyd}, LISA \cite{Robson:2018ifk, LISACosmologyWorkingGroup:2022jok}, Taiji \cite{Ruan:2018tsw, Wu:2018clg, Hu:2017mde, Gong:2021gvw}, and TianQin \cite{Liu:2020eko, Wang:2019ryf, TianQin:2015yph, Luo:2020bls, Gong:2021gvw}, among others, with their enhanced sensitivity and {substantially} extended detection ranges {and frequency coverage}, offer unprecedented opportunities to test GR and probe the nature of gravity with greater precision. In particular, space-based detectors extend the accessible frequency band to much lower frequencies (mHz to Hz), and this broader frequency coverage is a crucial factor for improving parameter estimation, for multi-band observations, and for enhancing sensitivity to potential deviations from GR.

To explore the nature of gravity with GWs, one important approach is to model the possible deviations from the standard GW waveforms in GR, which could arise from theories of gravity beyond GR. With specific derivations in the modified GW waveform from that in GR, one can obtain the constraints on the effects from GW data. However, different modifications to GR may induce different effects in both the generation and propagation of GW. Given a large number of modified theories beyond GR, one challenging task is to construct a unified framework for characterizing different effects so they can be directly tested with GW data in a model-independent way. Several parametrized frameworks have been proposed for this purpose \cite{Nishizawa:2017nef, Zhao:2019xmm, Zhu:2023rrx, Ezquiaga:2021ler, Tahura:2018zuq, Saltas:2014dha}. 

In this paper, we focus on probing GR through potential deviations from parity and Lorentz symmetries in the gravitational sector by using the observations of GWs. GR has several foundational symmetries, including diffeomorphism invariance, local Lorentz symmetry, and parity symmetry. Diffeomorphism invariance means the gravitational action of the theory is invariant under smooth, invertible coordinate transformations. In tetrad formulations of GR, the gravitational action is also invariant under the Lorentz transformation acting on the tangent space of an arbitrary point in the spacetime manifold. And the parity symmetry refers to invariance of the action under spatial inversion, so GR does not distinguish left- from right-handed spatial orientations unless parity-violating terms (e.g., a gravitational Chern–Simons term) are introduced.

While these symmetries are foundational to GR, their potential violations are motivated by theories that aim to bridge quantum physics and gravity. Parity symmetry is famously known to be broken in weak interactions \cite{Lee:1956qn, Wu:1957my}, and Lorentz symmetry has been validated to extraordinary precision in the context of particle physics experiments within the standard model \cite{Mattingly:2005re, Kostelecky:2008ts}. However, observational and experimental constraints on such violations in gravity have remained significantly less advanced. The violations of both the parity and Lorentz symmetry in gravity can lead to significant effects on the propagation of GWs on a homogeneous and isotropic cosmological background. With these new effects, one can compare them with GW data and obtain the corresponding constraints on the parity and Lorentz violations. To model the possible parity- and Lorentz-violating effects in GWs, we adopt a systematic parametric framework for characterizing possible derivations of GW propagation and modified waveforms with parity- and Lorentz-violating effects \cite{Zhu:2023rrx, Zhao:2019xmm}. It is shown in Refs.~\cite{Zhu:2023rrx, Zhao:2019xmm} that this parametrization provides a general framework for studying the GW propagation of possible modifications caused by various modified gravitational theories with Lorentz and parity violations. 

In this parametrization, parity- and Lorentz-violating effects are modeled in the amplitude and phase corrections to the GW waveforms, which are described by a series of deformation parameters. These corrections are frequency-dependent and thus distort the waveforms in GR. These new effects beyond GR can be divided into two classes: (1) parity-violating amplitude and velocity birefringences, and (2) the Lorentz-violating damping rate and dispersion of GWs. These effects are labeled by several different deformation parameters, with their frequency dependence described by factors $\beta_{\nu}$, $\beta_{\mu}$, $\beta_{\bar \nu}$, and $\beta_{\bar \mu}$. With these effects and the corresponding modified waveforms, one can obtain the constraints on the parity- and Lorentz-violating effects from GW data. This has enabled a lot of tests of parity and Lorentz symmetries by using GW signals detected by LVK \cite{LIGOScientific:2019fpa, LIGOScientific:2020tif, LIGOScientific:2021sio, Wang:2020cub, Wu:2021ndf, Gong:2021jgg, Zhao:2022pun, Wang:2021gqm, Haegel:2022ymk, Wang:2025fhw, Gong:2023ffb, Zhu:2022uoq, Niu:2022yhr, Zhao:2019szi, Wang:2025fhw}. Forecasts for constraining parity and Lorentz violation with future ground-based and space-based GW detectors have also been considered in \cite{Lin:2024pkr, Zhang:2024rel, Califano:2023aji, Mirshekari:2011yq, Zhang:2025kcw}. Note that parity- and Lorentz-violating effects can also be direction‑dependent when spatial rotation symmetry of gravity is broken, leading to anisotropic GW propagation \cite{Kostelecky:2016kfm}. Such anisotropic signatures can be tested by searching for sky‑location dependence in GW signals, and previous studies have constrained anisotropic coefficients (over various mass dimensions) using global‑fitting \cite{Wang:2021ctl, Haegel:2022ymk, Shao:2020shv} or maximum‑reach \cite{Niu:2022yhr, Wang:2025fhw, Gong:2023ffb, ONeal-Ault:2021uwu} approaches. In this work, we focus exclusively on isotropic deviations in GW propagation, and anisotropic effects are left for future investigation.

Most current analyses on the GW data from the LVK collaboration typically adopted two different ways: the single-parameter and multi-parameter approaches, see refs. \cite{LIGOScientific:2019fpa, LIGOScientific:2020tif, LIGOScientific:2021sio, Haegel:2022ymk, Zhu:2022uoq} for examples. In the commonly used single-parameter analysis, one parity- or Lorentz-violating parameter is varied at a time while all other deformation parameters are fixed to their GR values (i.e., zero). By contrast, multi-parameter analyses vary multiple deformation parameters simultaneously, thereby reflecting scenarios in which several deviations (or physical effects) may coexist, for instance, spatial covariant gravity provides such an example~\cite{Gao:2019liu}. However, they are computationally more demanding and, in general, less straightforward to map onto the wide variety of parity- and Lorentz-violating theories. In turn, single-parameter analyses are simpler to implement, and their constraints can often be directly associated with individual physical couplings. The resulting constraint on the chosen parameter can then be mapped to the corresponding coupling constants in specific modified theories of gravity (see, e.g., Ref.~\cite{Zhu:2022uoq}). This raises an important question: are the constraints derived from a {\em single-parameter analysis} robust under the assumption of ignorance regarding other deformation parameters? 

In general, one would expect that a simultaneous estimation of all relevant deformation parameters through a {\em multi-parameter analysis} would yield more robust and comprehensive constraints on parity and Lorentz violations in modified gravity theories. However, this approach demands an exhaustive exploration of the possible combinations of deformation parameters, which is computationally expensive and inefficient, given a large number of modified theories of gravity in the literature.

A similar issue has been investigated in the context of the parametrized inspiral test of GR within the parametrized post-Einsteinian framework \cite{Perkins:2022fhr}. This framework introduces fractional deformation parameters to modify the GW phase predicted by the post-Newtonian approximation to GR at each post-Newtonian order. Using a new prior of the deformation parameters that restricts the impact of the covariances of multi-parameters, it is shown in ref.~\cite{Perkins:2022fhr} that the single-parameter analysis of the parametrized inspiral waveform is quite robust compared to the multi-parameter analysis. Motivated by these findings, this paper examines the robustness of single-parameter analysis for constraining parity and Lorentz violations using the parametric framework described in \cite{Zhu:2023rrx, Zhao:2019xmm}. Specifically, we consider a set of multi-parameter GW waveforms incorporating parity- and Lorentz-violating effects and conduct a series of full Bayesian parameter estimation studies. These multi-parameter waveforms are well-motivated by several modified gravity theories. By comparing the constraints obtained from the multi-parameter analysis with those from the single-parameter analysis, we find that the two are generally consistent in magnitude. For some models, the inclusion of multiple parameters leads to slightly weaker (but still comparable) constraints.

This paper is organized as follows. In Sec.~II, we introduce the parametrization for characterizing the parity- and Lorentz-violating effects in GW propagation and the corresponding modified waveforms. In Sec.~III, we construct four specific multi-parameter waveform models, and in Sec.~IV, a brief introduction to Bayesian analysis and the settings of GW detectors used in our work are presented. In Sec.~V, the main results of constraining the parity- and Lorentz-violating effects are presented and discussed. We summarize the results and conclusions in Sec.~VI. 

Throughout this paper, the metric convention is chosen as $(-, +, +, +)$, and Greek indices $(\mu, \nu, \cdots)$ run over 0, 1, 2, 3, and Latin indices $(i, j, k, \cdots)$ run over 1, 2, 3.

\section{Modified waveforms with parity and Lorentz violations}\label{sec:2}
\renewcommand{\theequation}{2.\arabic{equation}} \setcounter{equation}{0}

In this section, we provide an introduction to the universal parametrization framework, which systematically describes deviations from gravitational wave (GW) propagation in GR \cite{Zhu:2023rrx, Zhao:2019xmm}. Most of the expressions and results used here can be found in \cite{Zhu:2023rrx, Zhao:2019xmm}. 

This framework is based on the parametrized equation of motion for the circular polarization modes of GWs \cite{Zhu:2023rrx, Zhao:2019xmm},
\begin{eqnarray}\label{eom_A}
 h''_A + (2+\bar \nu + \nu_A) \mathcal{H} h'_A + (1+\bar \mu+ \mu_A) k^2 h_A=0,\nb\\
\end{eqnarray}
where $A = {\rm R}$ or ${\rm L}$ represents the right- and left-handed circular polarization modes, respectively. Here, a prime denotes differentiation with respect to the conformal time $\tau$, $\mathcal{H} = a'/a$ is the conformal Hubble parameter with $a$ being the scale factor of the universe, and $k$ is the {comoving} wavenumber, {which is related to the GW frequency $f$ at the detector's frame by $ f =k/(2 \pi a_0)$ with the common convention $a_0=1$ at present}. In this parametrization, deviations from GR arising from theories beyond GR are fully characterized by four parameters: $\bar\nu$, $\bar\mu$, $\nu_A$, and $\mu_A$, each corresponding to distinct physical effects on GW propagation. These effects can be classified into three main categories. First, the frequency-independent modifications, governed by $\bar\nu$ and $\bar\mu$, alter the GW speed and introduce friction during propagation. Second, parity-violating effects, described by $\nu_A$ and $\mu_A$, lead to amplitude birefringence and velocity birefringence between the two polarization modes. Finally, frequency-dependent effects, associated with $\bar\nu$ and $\bar\mu$, result in frequency-dependent damping and nonlinear dispersion. While the first case does not relate to the symmetry breaking, the rest two cases correspond to the effects of parity and Lorentz violation in gravity, respectively. The specific forms of these parameters in various modified gravity theories, such as ${\cal H}\bar \nu$, $\bar \mu$, ${\cal H}\nu_A$, and $\mu_A$, are summarized in Table I of Ref.~\cite{Zhu:2023rrx}. In this paper, we only focus on the latter two cases. 

The parameters $\nu_A$ and $\mu_A$ label the gravitational parity-violating effects. The parameter $\mu_A$ induces velocity birefringence, leading to different velocities of left- and right-hand circular polarizations of GWs, so the arrival times of them are different. The parameter $\nu_A$, on the other hand, induces amplitude birefringence, leading to different damping rates of left- and right-hand circular polarizations of GWs, so the amplitude of the left-hand mode increases (or decreases) during its propagation, while the amplitude of the right-hand mode decreases (or increases). For a large number of parity-violating theories, $\nu_A$ and $\mu_A$ are frequency-dependent. Thus, one can further parametrize $\nu_A$ and $\mu_A$ as \cite{Zhao:2019xmm}
\begin{eqnarray}
\mathcal{H} \nu_{\mathrm{A}} &=&\left[\rho_{\mathrm{A}} \alpha_{\nu}(\tau)\left(k / a M_{\mathrm{PV}}\right)^{\beta_{\nu}}\right]^{\prime}, \\
\mu_{\mathrm{A}}&=&\rho_{\mathrm{A}} \alpha_{\mu}(\tau)\left(k / a M_{\mathrm{PV}}\right)^{\beta_{\mu}},
\end{eqnarray}
where $\beta_\nu$, $\beta_\mu$ are arbitrary numbers, $\alpha_\nu$, $\alpha_\mu$ are arbitrary functions of time, and $M_{\rm PV}$ denotes the energy scale of the parity violation. Here $\rho_{\rm R}=1$ and $\rho_{\rm L}=-1$ respectively.

Violations of Lorentz symmetry can lead to nonzero and frequency-dependent $\bar \nu$ and $\bar \mu$. The parameter $\bar \mu$ induces frequency-dependent friction in the propagation equation of GWs, while $\bar \mu$ modifies the conventional linear dispersion relation of GWs to nonlinear ones. Considering both $\bar \nu$ and $\bar \mu$ are frequency-dependent, one can parametrize them as
 \begin{eqnarray}
\mathcal{H} \bar{\nu} &=&\left[\alpha_{\bar{\nu}}(\tau)\left(k / a M_{\mathrm{LV}}\right)^{\beta_{\bar{\nu}}}\right]', \\
 \bar{\mu}&=&\alpha_{\bar{\mu}}(\tau)\left(k / a M_{\mathrm{LV}}\right)^{\beta_{\bar{\mu}}}, 
 \end{eqnarray}
where $\beta_{\bar \nu}$, $\beta_{\bar \mu}$ are arbitrary numbers, $\alpha_{\bar \nu}$, $\alpha_{\bar \mu}$ are arbitrary functions of time, and $M_{\rm LV}$ denotes the energy scale of Lorentz violation. 

With the above parametrization, one can derive their explicit GW waveforms by solving the equation of motion (\ref{eom_A}). It is shown \cite{Zhao:2019xmm} that the amplitude and phase modifications to the GR-based waveform due to the parity- and Lorentz-violating effects can be written as
\begin{eqnarray}
 \tilde h_A(f) = \tilde h_A^{\rm GR}(f) e^{ \rho_A \delta h_1 +\delta h_2} e^{i  (\rho_A \delta \Psi_1 + \delta \Psi_2)}.\label{waveforms}
 \end{eqnarray}
 where $ \tilde h_A^{\rm GR} $ is the corresponding GR-waveform, and its explicit form can be found in the previous works \cite{Zhao:2019xmm}. The amplitude correction $\delta h_1 = A_{\nu} (\pi f)^{\beta_\nu}$ and $\delta h_2 = -A_{\bar \nu} (\pi f)^{\beta_{\bar \nu}}$ are caused by the parameters $\nu_A$ and $\bar{\nu}$, respectively, while the phase corrections $\delta \Psi_1$ and $\delta \Psi_2$ are induced by the parameters $\mu_A$ and $\bar{\mu}$ respectively. Their explicit forms are given by
 \begin{eqnarray}
 \delta h_1 &=& A_{\nu} (\pi f)^{\beta_\nu}, \\
 \delta \Psi_1 &=& \begin{cases}
                  A_{\mu} (\pi f)^{\beta_\mu+1}, & \beta_\mu \neq -1, \\
                  A_{\mu} \ln u, & \beta_\mu =- 1, 
                  \end{cases} 
\end{eqnarray}
for parity-violating effects and
\begin{eqnarray}
 \delta h_2 &= & -A_{\bar \nu} (\pi f)^{\beta_{\bar{\nu}}}, \\
\delta \Psi_2 &=&  \begin{cases}
                  A_{\bar \mu} (\pi f)^{\beta_{\bar \mu}+1}, & \beta_{\bar \mu} \neq -1, \\
                  A_{\bar \mu} \ln u, & \beta_{\bar \mu} =- 1,
                  \end{cases}
\end{eqnarray}
for Lorentz-violating effects, with the four coefficients $A_{\nu}$, $A_{\bar \nu}$, $A_{\mu}$, and $A_{\bar \mu}$ are given by
\begin{eqnarray}
    A_{\nu} &=& \frac{1}{2} \left(\frac{2}{M_{\rm PV}}\right)^{\beta_\nu}\Big[\alpha_\nu(\tau_0) - \alpha_\nu(\tau_e) (1+z)^{\beta_\nu}\Big], \label{Anu2.11} \nb\\
  &&  \; \\
A_{\mu} &=& \frac{(2/M_{\rm PV})^{\beta_\mu}}{\Theta(\beta_\mu+1)}  \int_{t_e}^{t_0} \frac{\alpha_\mu}{a^{\beta_\mu+1}}dt, \label{Amu}\\
A_{\bar \nu} &=& \frac{1}{2} \left(\frac{2}{M_{\rm LV}}\right)^{\beta_{\bar \nu}}\Big[\alpha_{\bar \nu}(\tau_0) - \alpha_{\bar \nu}(\tau_e) (1+z)^{\beta_{\bar \nu}}\Big], \label{Abnu}\nb\\
  &&  \; \\
A_{\bar \mu} &=&  \frac{(2/M_{\rm LV})^{\beta_{\bar \mu}}}{\Theta(\beta_{\bar \mu}+1)} \int_{t_e}^{t_0} \frac{\alpha_{\bar \mu}}{a^{\beta_{\bar \mu}+1}}dt. \label{Abmu2.14}
\end{eqnarray}
 Here $t_e$ ($t_0$) is the emitted (arrival) time for a GW event, $z=1/a(t_e) -1 $ is the redshift, $f$ is the GW frequency at the detector, and $u=\pi {\cal M} f $ with ${\cal M}$ being the measured chirp mass of the binary system. {The chirp mass is defined in terms of the component masses $m_1$ and $m_2$ as $\mathcal{M} \equiv \left(m_1 m_2\right)^{3 / 5} /\left(m_1+m_2\right)^{1 / 5}$.} The function $\Theta(1+x)=1+x$ for $x\neq -1$ and $\Theta(1+x)=1$ for $x= -1$. Throughout this work, we assume the four functions, $\alpha_\nu$, $\alpha_{\mu}$, $\alpha_{\bar \nu}$, and $\alpha_{\bar \mu}$ vary slowly over the redshift range we probed in our analysis (up to $z \lesssim 1$). {This choice is motivated by the expectation that any evolution driven by the cosmological background is slow over this redshift range, so the constant approximation is adequate at current data precision. Relaxing this assumption in favor of an explicit time-dependent parameterization is left for future work when higher-redshift samples are available.}
 In this work, the cosmology parameter we adopt are $\Omega_m=0.315$, $\Omega_{\Lambda}=0.685$, and $H_0=67.4\; {\rm km}\;{\rm s}^{-1}\; {\rm Mpc}^{-1}$ \cite{Planck:2018vyg}.

{At last, we would like to mention that the modified waveforms presented in the above include only frequency-dependent propagation corrections. As shown in ref.~\cite{Zhu:2023rrx}, some Lorentz- and parity-violating models predict frequency-independent friction or speed modification as well. The frequency-independent speed modification has already tightly constrained, thus it is expected to have only negligible effects on the waveforms. And for frequency-independent friction change, since it results in a frequency-independent change, it is exactly degenerate with the GR luminosity distance. Adding a frequency-independent amplitude parameter would only lead to an overall amplitude scaling. For these reasons, we only consider the frequency-dependent corrections in our later analysis.}

\section{Specific models with multi-parameter waveforms}\label{sec:3}
\renewcommand{\theequation}{3.\arabic{equation}} \setcounter{equation}{0}

As an extension of previous work \cite{Zhu:2023rrx}, we investigate several selected combinations of deformation parameters within the framework of modified gravity theories, specifically, $(\beta_\nu, \beta_\mu) = (1, -1),\ (1, 1)$, $(\beta_{\bar{\nu}}, \beta_{\bar{\mu}}) = (2, 2)$, {$(\beta_{\bar{\mu}}^{(2)},\beta_{\bar{\mu}}^{(4)})=\ (2, 4)$, and $(\beta_\nu, \beta_\mu, \beta_{\bar{\nu}}, \beta_{\bar{\mu}})=\ (1, 1, 2, 2)$,} to assess whether a multi-parameter analysis yields more robust results. These models are well-motivated by several specific modified gravities with parity and Lorentz violations \footnote{Note that here we restrict the values of $\beta_\nu$, $\beta_\mu$, $\beta_{\bar \nu}$, and $\beta_{\bar \mu}$ to the discrete integer values; different integers correspond to distinct theoretical models rather than to a continuous family.}. The corresponding modified waveforms for each of these four distinct cases are presented separately below.

\subsection{Model 1: $\beta_\nu=1$ and $\beta_\mu=-1$}

The case with  $\beta_\nu=1$ and $\beta_\mu=-1$ can arise from the parity-violating Kalb-Ramond gravity \cite{Manton:2024hyc, Altschul:2009ae}. It can also appear in the parity-violating scalar-nonmetricity gravity as a specific case \cite{Chen:2022wtz, Conroy:2019ibo, Li:2022vtn}. This case leads to both the amplitude and velocity briefringences of GWs. The  modified waveforms for this case can be written as
\begin{eqnarray}
 \tilde h_A(f) = \tilde h_A^{\rm GR}(f) e^{ \rho_A \delta h_1} e^{i \rho_A \delta \Psi_1 },\label{waveforms}
\end{eqnarray}
with the amplitude and phase corrections being given by
\begin{eqnarray}
\delta h_1 = A_{\nu} \pi f, \\
\delta \Psi_1 = A_\mu \ln u.
\end{eqnarray}
Here $A_\nu$ and $A_\mu$ are given by 
\begin{eqnarray}
A_\nu &=& 
- \frac{\alpha_\nu}{M_{\rm PV}} z,\\
A_\mu &=& \frac{\alpha_\mu M_{\mathrm{PV}}}{2} \int_0^z \frac{\left(1+z^{\prime}\right)^{-1}}{H_0 \sqrt{\Omega_m\left(1+z^{\prime}\right)^3+\Omega_{\Lambda}}} d z^{\prime}.
\end{eqnarray}
Note that in the above, we have treated $\alpha_\nu$ and $\alpha_\mu$ as constants. In parity-violating Kalb-Ramond gravity \cite{Manton:2024hyc, Altschul:2009ae}, the coefficients $\alpha_\nu$ and $\alpha_{\mu}$ are not independent; they are related to the same coupling constant of the theory. And in parity-violating scalar-nonmetricity gravity \cite{Chen:2022wtz, Conroy:2019ibo, Li:2022vtn}, these two coefficients relate to different combinations of coupling constants of the theory and are independent of each other.

\begin{figure}
  \centering
  \includegraphics[width=0.47\textwidth]{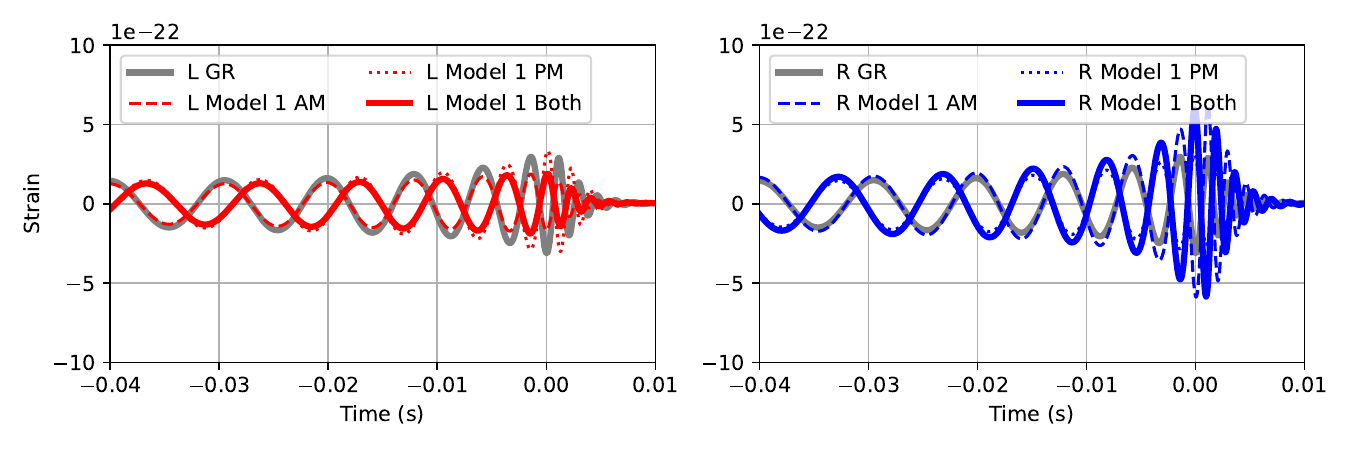}
  \includegraphics[width=0.47\textwidth]{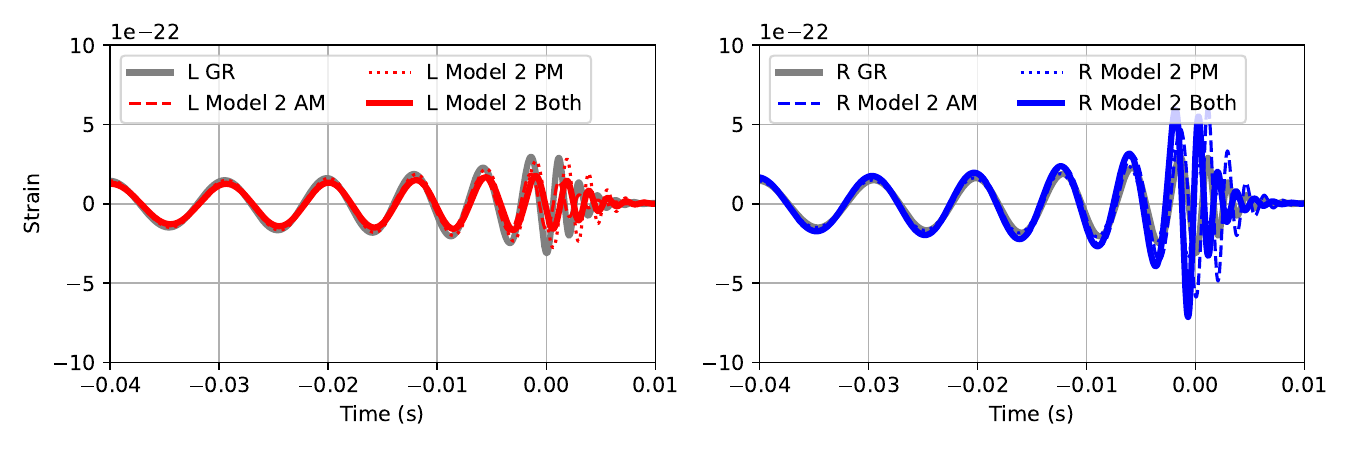}
  \includegraphics[width=0.47\textwidth]{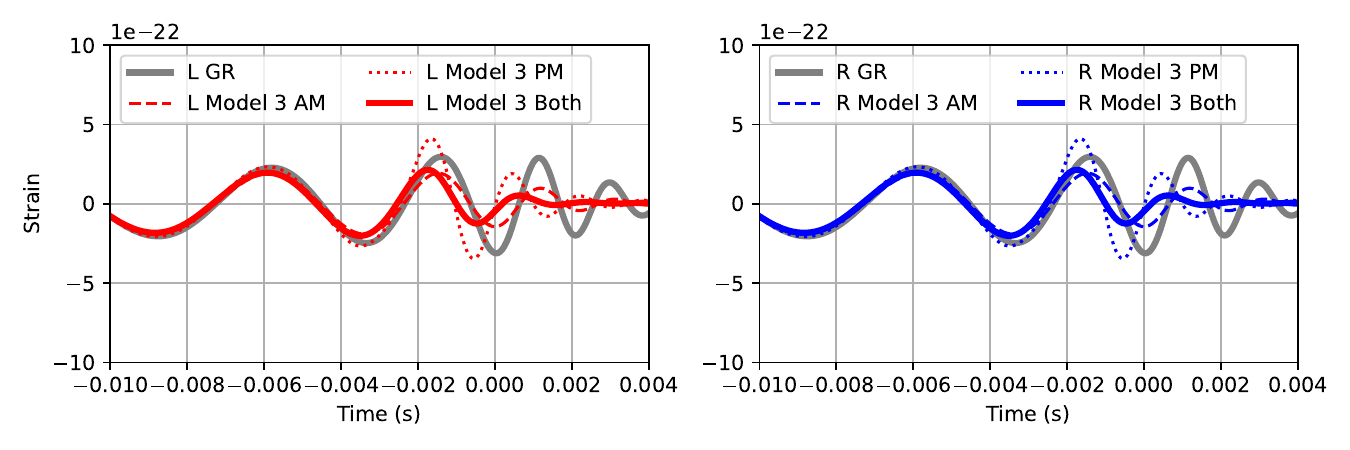}
  \includegraphics[width=0.47\textwidth]{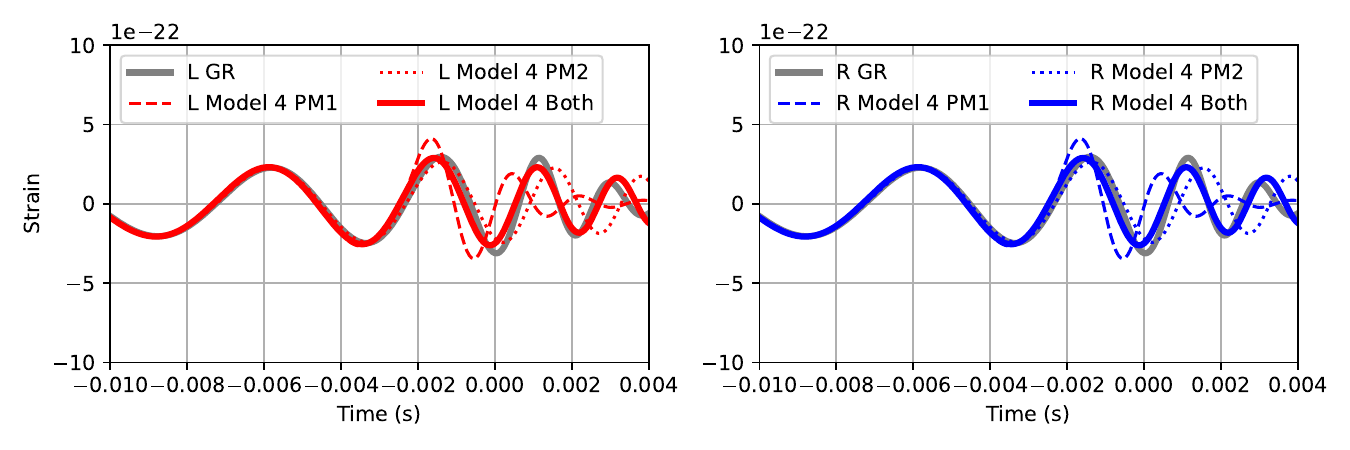}
  \caption{{Demonstration of the impact of parity and Lorentz violations on the left-circular polarization components (left column) and the right-circular polarization components (right column) of gravitational waves. Grey solid curves are the case of standard GR; the labels `AM' and `PM' stand for amplitude and phase modifications, respectively, and `Both' represents the two modifications in a model are all applied. The zero point of time is taken as the moment when the strain of the standard GR waveform reaches its maximum.}}
  \label{fig:WFs}
\end{figure}

\subsection{Model 2: $\beta_\nu=1$ and $\beta_\mu=1$}

The case with  $\beta_\nu=1$ and $\beta_\mu=1$ can arise from the Palatini Chern-Simons gravity \cite{Sulantay:2022sag} and Chiral-scalar-tensor theory \cite{Crisostomi:2017ugk, Qiao:2019wsh, Nishizawa:2018srh}. It can also appear in the parity-violating scalar-nonmetricity gravity \cite{Chen:2022wtz, Conroy:2019ibo, Li:2022vtn} and spatial covariant gravity \cite{Gao:2019liu, Zhu:2022dfq} as a specific case. This case also leads to both the amplitude and velocity briefringences of GWs. The  modified waveforms for this case can be written as
\begin{eqnarray}
 \tilde h_A(f) = \tilde h_A^{\rm GR}(f) e^{ \rho_A \delta h_1} e^{i \rho_A \delta \Psi_1 },\label{waveforms}
\end{eqnarray}
with the amplitude and phase corrections being given by
\begin{eqnarray}
\delta h_1 = A_{\nu} \pi f, \\
\delta \Psi_1 = A_\mu (\pi f)^2.
\end{eqnarray}
Here $A_\nu$ and $A_\mu$ are given by 
\begin{eqnarray}
A_\nu &=& 
- \frac{\alpha_\nu}{M_{\rm PV}} z,\\
A_\mu &=& \frac{\alpha_\mu}{M_{\mathrm{PV}}} \int_0^z \frac{1+z^{\prime}}{H_0 \sqrt{\Omega_m\left(1+z^{\prime}\right)^3+\Omega_{\Lambda}}} d z^{\prime}.
\end{eqnarray}
Note that in the above we have treated $\alpha_\nu$ and $\alpha_\mu$ as constants.

\subsection{Model 3: $\beta_{\bar \nu}=2$ and $\beta_{\bar \mu}=2$}

The case with $\beta_{\bar \nu}=2$ and $\beta_{\bar \mu}=2$ can arise from the Horava-Lifshitz gravity with mixed derivative coupling \cite{Colombo:2014lta} and the spatial covariant gravity \cite{Gao:2019liu, Zhu:2022dfq} when the sixth-order spatial derivative terms are absent. This case leads to a Lorentz-violating damping rate and a nonlinear dispersion relation of GWs. The  modified waveforms for this case can be written as
\begin{eqnarray}
 \tilde h_A(f) = \tilde h_A^{\rm GR}(f) e^{\delta h_2}e^{i \delta \Psi_2},\label{waveforms}
\end{eqnarray}
with the amplitude and phase corrections being given by
\begin{eqnarray}
\delta h_2 = - A_{\bar \nu} (\pi f)^2, \\
\delta \Psi_2 = A_{\bar \mu} (\pi f)^3.
\end{eqnarray}
Here $A_{\bar \nu} $ and $A_{\bar \mu}$ are given by 
\begin{eqnarray}
A_{\bar \nu}  &=& \frac{2}{M^2_{\mathrm{LV}}} \Big[\alpha_{\bar \nu} - \alpha_{\bar \nu} (1+z)^2\Big],\\
A_{\bar \mu} &=& \frac{8}{3}\frac{ \alpha_{\bar \mu}}{M^2_{\mathrm{LV}}} \int_0^z \frac{(1+z^{\prime})^2}{H_0 \sqrt{\Omega_m\left(1+z^{\prime}\right)^3+\Omega_{\Lambda}}} d z^{\prime}.\nb\\
\end{eqnarray}
Note that in the above we have treated $\alpha_{\bar \nu}$ and $\alpha_{\bar \mu}$ as constants. 

\subsection{Model 4: $\beta_{\bar{\mu}}^{(2)}=2$ and $\beta_{\bar{\mu}}^{(4)}=4$}

The case with $\beta_{\bar{\mu}}^{(2)}=2$ and $\beta_{\bar{\mu}}^{(4)}=4$ can arise from the Horava-Lifshitz gravity \cite{Horava:2009uw, Blas:2009qj, Blas:2010hb, Zhu:2011yu, Zhu:2011xe}. It can also appear in the spatial covariant gravity \cite{Gao:2019liu, Zhu:2022dfq} and diffeomorphism- and Lorentz-violating linearized gravity \cite{Kostelecky:2017zob, Mewes:2019dhj} as a specific case. This case leads to a nonlinear dispersion relation of GWs with different frequency dependencies. The  modified waveforms for this case can be written as
\begin{eqnarray}
 \tilde h_A(f) = \tilde h_A^{\rm GR}(f) e^{i ( \delta \Psi_2^{(2)} + \delta \Psi_2^{(4)})},\label{waveforms}
\end{eqnarray}
with phase corrections being given by
\begin{eqnarray}
\delta \Psi_2^{(2)} = A_{\bar \mu}^{(2)} (\pi f)^3,\\
\delta \Psi_2^{(4)} = A_{\bar \mu}^{(4)} (\pi f)^5.
\end{eqnarray}
Here $A_{\bar \mu}^{(2)}$ and $A_{\bar \mu}^{(4)}$ are given by 
\begin{eqnarray}
A_{\bar \mu}^{(2)} &=& \frac{8}{3}\frac{ \alpha_{\bar \mu}^{(2)}}{M^2_{\mathrm{LV}}} \int_0^z \frac{(1+z^{\prime})^2}{H_0 \sqrt{\Omega_m\left(1+z^{\prime}\right)^3+\Omega_{\Lambda}}} d z^{\prime}. \\
A_{\bar \mu}^{(4)} &=& \frac{16}{5}\frac{ \alpha_{\bar \mu}^{(4)}}{M^4_{\mathrm LV}} \int_0^z \frac{(1+z^{\prime})^4}{H_0 \sqrt{\Omega_m\left(1+z^{\prime}\right)^3+\Omega_{\Lambda}}} d z^{\prime}.\nb\\
\end{eqnarray}
Note that in the above we have treated $\alpha_{\bar \mu}^{(2)}$ and $\alpha_{\bar \mu}^{(4)}$ as constants. 

{As a summary of the models described above, we demonstrate the impact of parity and Lorentz violations in Figure.\ref{fig:WFs}. The figure compares the modified waveforms (left- and right-circular polarization components) with the predictions of GR for the four models considered in this study. The GR waveform (grey solid curves) is generated from a BBH system with component masses of $15M_\odot+15M_\odot$, zero spins, an inclination angle of $\pi/2$, and a luminosity distance of $400\ {\rm Mpc}$. The dashed and dotted curves show the modified waveforms when only the amplitude correction and only the phase correction are included, respectively. The red (blue) solid curves correspond to the waveforms with two modifications as described in each model for the left (right) handed circular polarization mode. As shown in the figure, Model 1 and Model 2 incorporate parity-violating effects, which induce distinct amplitude and phase modifications in the two modes. In contrast, Model 3 and Model 4 involve only Lorentz violations and therefore produce identical modifications in both polarization modes. Notably, Model 4 includes two phase correction terms with different frequency dependencies, and for certain parameter choices, their effects on the waveform can partially cancel, as illustrated in the last panel of the figure. In addition, as a supplement to the above four models,  we also describe a four-parameter model in Appendix B that explicitly accounts for both Lorentz- and parity-violating effects concurrently. This model integrates the key features of both phenomena, allowing us to evaluate their combined impact on the parameter estimation. }

\section{Bayesian inferences on the Modified waveforms with GW signals}
\renewcommand{\theequation}{4.\arabic{equation}} \setcounter{equation}{0}

Bayesian inference has been widely used in gravitational-wave astronomy, and it is an effective tool that allows us to test and place constraints on modified theories of gravity. {Under this framework, the primary goal is to construct a posterior distribution $P(\vec{\theta} \mid d, H)$, where $d$ is the data and $\vec{\theta}$ is the set of parameters in model $H$. In the case of compact binary coalescence, $d$ is the strain data from a network of gravitational-wave detectors\cite{Thrane:2018qnx}. To describe the merger of a BBH system, $\vec{\theta}$ includes 15 parameters In standard GR; In the models described in Sec.\ref{sec:3}, $\vec{\theta}$ also includes additional parameters ($A_\nu$, $A_\mu$, $A_{\bar{\nu}}$, and $A_{\bar{\mu}}$) describing the modifications of amplitude and/or phase on the waveform. The posterior distribution of $\vec{\theta}$ can be written as:}
\begin{eqnarray}
 P(\vec{\theta} \mid d, H) = \frac{\mathcal{L}(d \mid \vec{\theta}, H) \, \pi (\vec{\theta} \mid H)}{\mathcal{Z} (d \mid H)},\label{Baysian}
\end{eqnarray}
where $\mathcal{L}(d \mid \vec{\theta}, H)$ is the likelihood of observing the data given a specific set of $\vec{\theta}$, $\pi (\vec{\theta} \mid H)$ is the prior distribution, and $\mathcal{Z} (d \mid H)$ is known as the evidence that is defined by the integral
\begin{eqnarray}
\mathcal{Z} (d \mid H) \equiv  \int d\vec{\theta} \, \mathcal{L} (d \mid \vec{\theta}, H) \, \pi(\vec{\theta} \mid H).\label{evidence}
\end{eqnarray}

{To efficiently obtain the high-dimensional posterior distribution in E.q.(\ref{Baysian}), we employ nested sampling algorithm \cite{2021PhRvD.103j3006W} implemented in the Bilby library \cite{Romero-Shaw:2020owr, Ashton:2018jfp}. The algorithm could generate samples drawn from the posterior distribution, so that we could access the constraints on parameters by performing statistics on the posterior samples. The likelihood used in E.q.(\ref{Baysian}) is computed in the frequency domain assuming stationary Gaussian noise, with the noise power spectral density estimated by \cite{KAGRA:2021vkt} for each event.}

The modified waveforms are constructed based on the approaches described in \cite{Zhu:2023rrx}. In brief, for a particular set of $\vec{\theta}$, we first generate the GR waveform in the frequency domain using the IMRPhenomXPHM template \citep{Pratten2021} implemented in the LALSuite \citep{LALSuite}, and then parity- or Lorentz-violating effects are added to the waveform according to Eq.(\ref{waveforms}). The circular polarization modes and linear polarization modes are converted to each other using the relations 
\begin{eqnarray}
\tilde{h}_{+}&=&\frac{\tilde{h}_{\mathrm{L}}+\tilde{h}_{\mathrm{R}}}{\sqrt{2}}, \\
 \tilde{h}_{\times}&=&\frac{\tilde{h}_{\mathrm{L}}-\tilde{h}_{\mathrm{R}}}{\sqrt{2} i}.
\end{eqnarray}
{Ref.~\cite{Zhu:2023rrx} has performed a comprehensive analysis of O1–O3 events for several parity- and Lorentz-violating cases; building on the work of \cite{Zhu:2023rrx}, our selection criteria of events can be summarized as follows: (1). When an equivalent single-parameter analysis exists in \cite{Zhu:2023rrx} (e.g, both the $A_{\nu} = 0$ and $A_{\mu} = 0$ cases in Model 1), we selected four events from the six that gave the tightest constraints in \cite{Zhu:2023rrx}. (2). When no equivalent single-parameter result is available in \cite{Zhu:2023rrx}, we used the same set of events as for the other case within the same model. For example, in Model 2, \cite{Zhu:2023rrx} only reported constraints for the $A_{\nu} = 0$ case. Therefore, we adopted the same events as in the $A_{\mu} = 0$ case; (3). We only consider BBH systems in this study, and for the $A_{{\bar \mu},1} = 0$ case in Model 4, we adopt two events (GW190720\_000836 and GW200225\_060421) from the $A_{{\bar \mu},2} = 0$ case to maintain a consistent number of sources across the analysis. As a result, eight events are selected to constrain Lorentz violation and another eight to constrain both violations, and we summarize them in Table \ref{tab:tab1}. For the four-parameter model described in Appendix B, which includes both the Lorentz and parity-violating effects, we adopt the loudest GW event, GW250114\_082203, for both the multi- and single-parameter analysis.}

For each event, the publicly available strain data sampled at 4096 Hz are used. To balance noise suppression, computational efficiency, and the characteristics of the physical signal, we set the minimum and maximum frequencies in the analysis to 20 {Hz} and 2048 {Hz} \footnote{Most of our selected events have chirp masses $>10M_\odot$, and the 20 {Hz} - 2048 {Hz} band fully encompasses their late inspiral, merger, and ringdown stages. The results remain unchanged when using an extended band of 20 {Hz}-4096 {Hz} to analyze the lightest event GW190924\_021846, respectively.}.

{For the deformation parameters $A_\nu$, $A_\mu$, $A_{\bar{\nu}}$, and $A_{\bar{\mu}}$ we adopt two prior strategies to perform distinct tests of each model described above:}

\begin{enumerate}
\item \textbf{Single-parameter analysis}: {In each model (with specified beta values), one of the two relevant deformation parameters is set to zero, while the other is assigned a Uniform prior in the inference. For example, in one single-parameter analysis for Model 1 with $\beta_\nu=1$ and $\beta_\mu=-1$, we fix $A_\nu = 0$ and constrain $M_{\mathrm{PV}}$ from $A_\mu$.} 

\item \textbf{Multi-parameter analysis}: {Both deformation parameters are assigned Uniform priors and are constrained in the inference. This corresponds to the full parameter space of the model, and we can compare the results with those of the single-parameter analysis to examine the possible degeneracies between parameters and how they affect the overall constraints.}
\end{enumerate}
The comparison between these two analyses reveals whether the constraints are dominated by one effect or require the combined contribution of both.
 For other GR parameters, we use identical configurations of priors as adopted in \cite{Zhu:2023rrx}. We employed a uniform prior on spin magnitude and an isotropic prior on spin orientation, as used in ref.~\cite{KAGRA:2021vkt}. For the chirp mass, we adopt uniform distributions (identical to the configurations in our previous paper \cite{Zhu:2023rrx}) with boundaries narrower than those in \cite{KAGRA:2021vkt} as prior for faster convergence. We have checked that using broader priors does not lead to notable changes in our results. We use power-Law priors (with index $\alpha=2$) as priors for the luminosity distance. The specific prior distributions for the mass ratio, chirp mass, luminosity distance, and the non-GR parameters used in each event are provided in the Table \ref{tab:tab3} in Appendix A.

\begin{table}[htbp]
\begin{ruledtabular}
\centering
\caption{\label{tab:tab1}Binary black hole events used for testing parity violation and Lorentz violation.}
\begin{tabular}{cc}
Events for Parity violation & Events for Lorentz violation \\
\hline
GW190727\_060333 & GW190512\_180714 \\
GW200112\_155838 & GW190707\_093326 \\
GW190513\_205428 & GW190708\_232457 \\
GW200311\_115853 & GW191215\_223052 \\
GW200128\_022011 & GW190720\_000836 \\
GW190910\_112807 & GW200225\_060421 \\
GW190915\_235702 &GW190924\_021846 \\
GW190731\_140936 & GW190412 \\
GW250114\_082203 & GW250114\_082203
\end{tabular}
\end{ruledtabular}
\end{table}

 \section{Results}
  \renewcommand{\theequation}{5.\arabic{equation}} \setcounter{equation}{0}

{Following \cite{Zhu:2023rrx}, the posterior samples (including the standard GR parameters and corresponding deformation parameters for each model) obtained from Bayesian inference are converted to that of $M_{\mathrm{PV}}^{-\beta_\nu}, M_{\mathrm{PV}}^{-\beta_\mu}$, $M_{\mathrm{LV}}^{-\beta_{\bar{\nu}}}, M_{\mathrm{LV}}^{-\beta_{\bar{\mu}}}$ via the mapping relations given in E.q.(\ref{Anu2.11})--(\ref{Abmu2.14}).}

\begin{figure*}
    \includegraphics[width=0.45\linewidth]{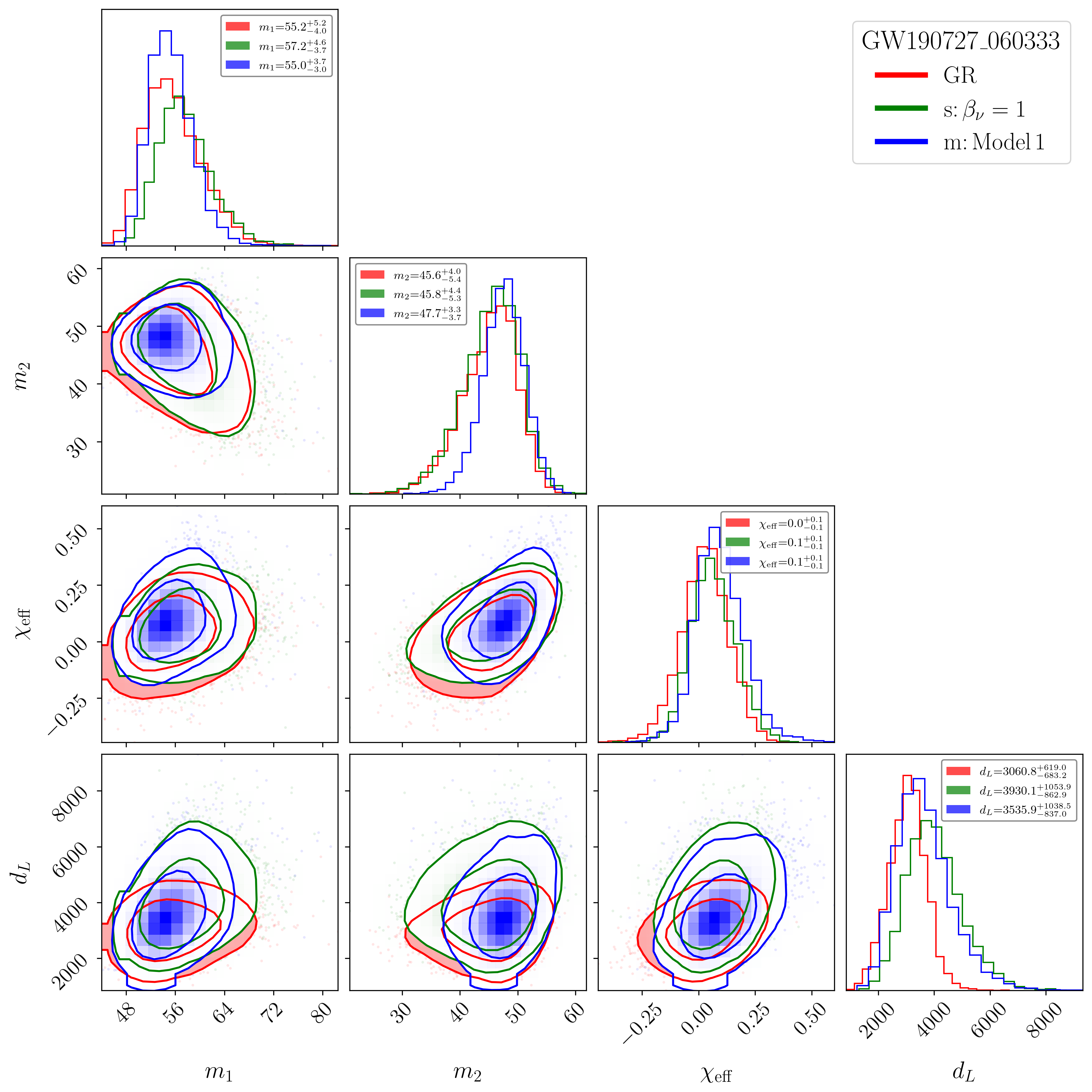}
    \includegraphics[width=0.45\linewidth]{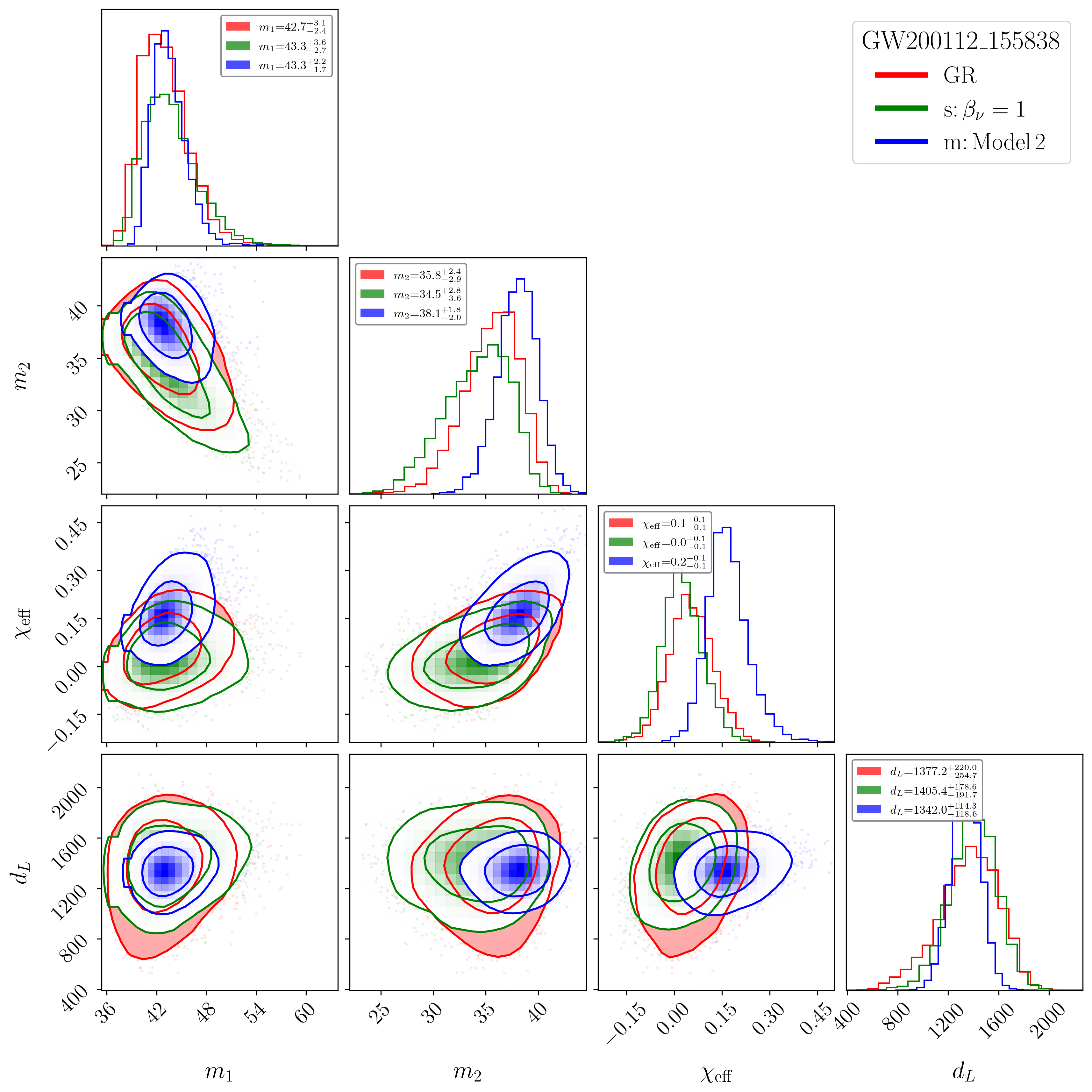}
    \includegraphics[width=0.45\linewidth]{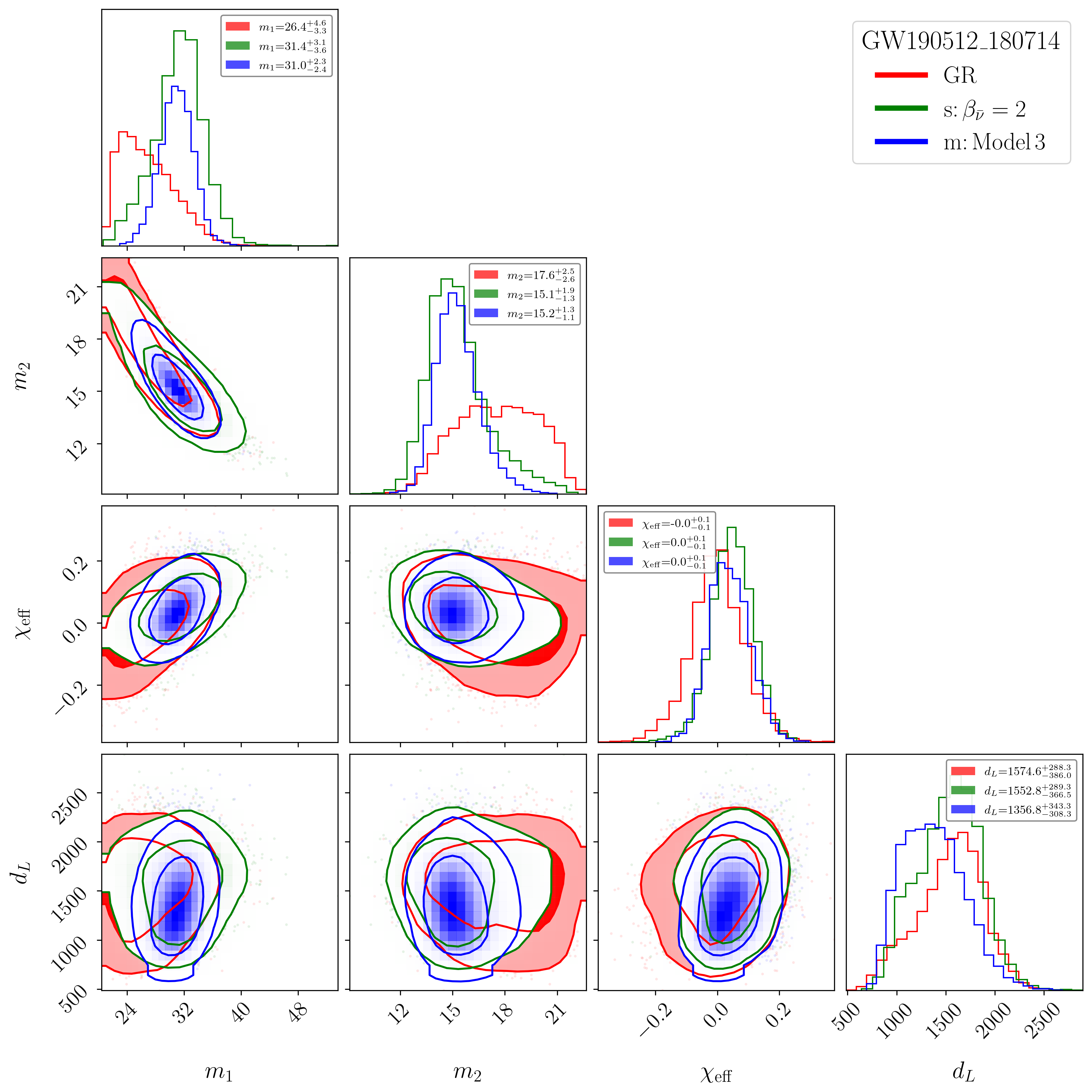}
    \includegraphics[width=0.45\linewidth]{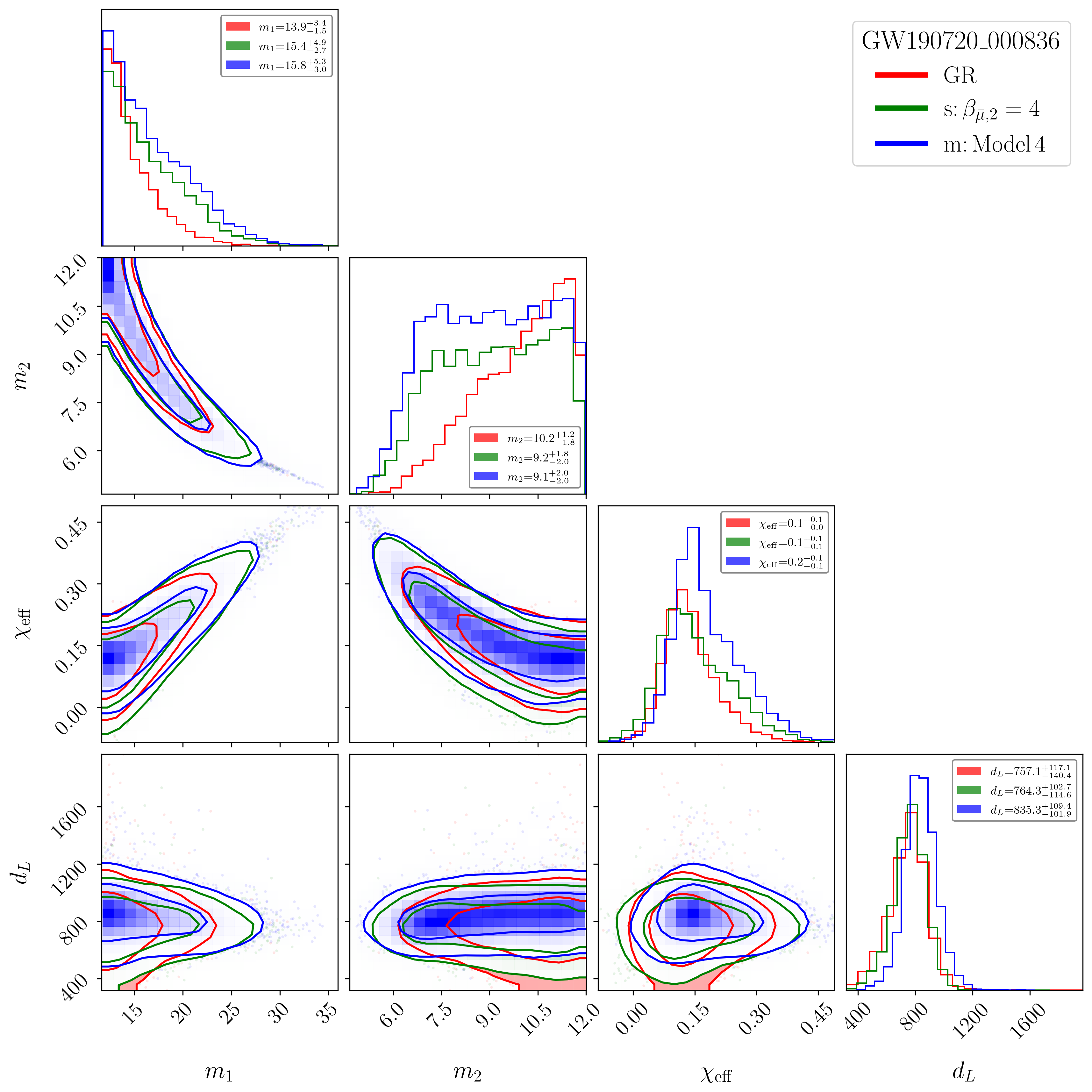}
    \caption{
{Corner plots of posterior distributions for one gravitational-wave event chosen for each of the four non-GR models. Shaded regions show the 50\%, 68\%, and 90\% credible intervals. Legend: GR—GR-based analysis; s—single-parameter non-GR analysis; m—two-parameter non-GR analysis. For each non-GR model, source-parameter posteriors from GR, s, and m analyses are compared for the selected event.}
}

 \label{comparison}
\end{figure*}

{For each model, we choose four events from the six that yield the strongest constraints in \cite{Zhu:2023rrx}. In Model 2, since constraints for $\beta_\mu=2$ are not reported in that study, the same event set is used for both the $\beta_\nu=1$ and $\beta_\mu=1$ cases. Likewise, for the $\beta_\mu=4$ case in Model 4, we exclude neutron-star events and adopt the same events employed for the $\beta_\mu=2$ constraint.} In Figure~\ref{comparison}, we present the corner plots of the posterior distributions for the source parameters of representative GW events, with one event selected for each of the four non-GR models. For each selected event, we compare the posteriors obtained using GR-based waveforms with those from both multi-parameter (labeled `m') and single-parameter (labeled `s') analyses of the four non-GR models. The shaded contours denote the 50\%, 68\%, and 90\% credible regions. {Here, $\chi_{\text {eff }}$ represents the effective spin parameter, defined as $\chi_{\text {eff }}=\left(m_1 \chi_{1 z}+m_2 \chi_{2 z}\right) /\left(m_1+m_2\right)$, where $m_{1,2}$ are the component masses and $\chi_{1 z, 2 z}$ are the spin projections along the orbital angular momentum.} Across all cases, the posterior distributions inferred with the non-GR waveforms are consistent with those obtained using GR-based waveforms.

\begin{figure}
  \centering
  \includegraphics[width=0.235\textwidth]{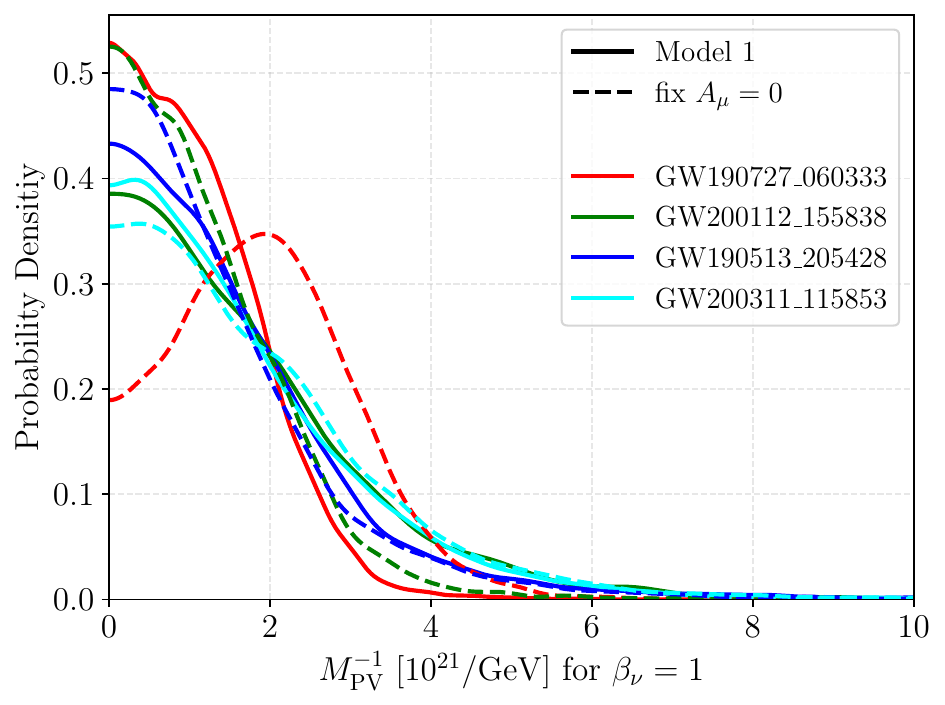}
  \includegraphics[width=0.235\textwidth]{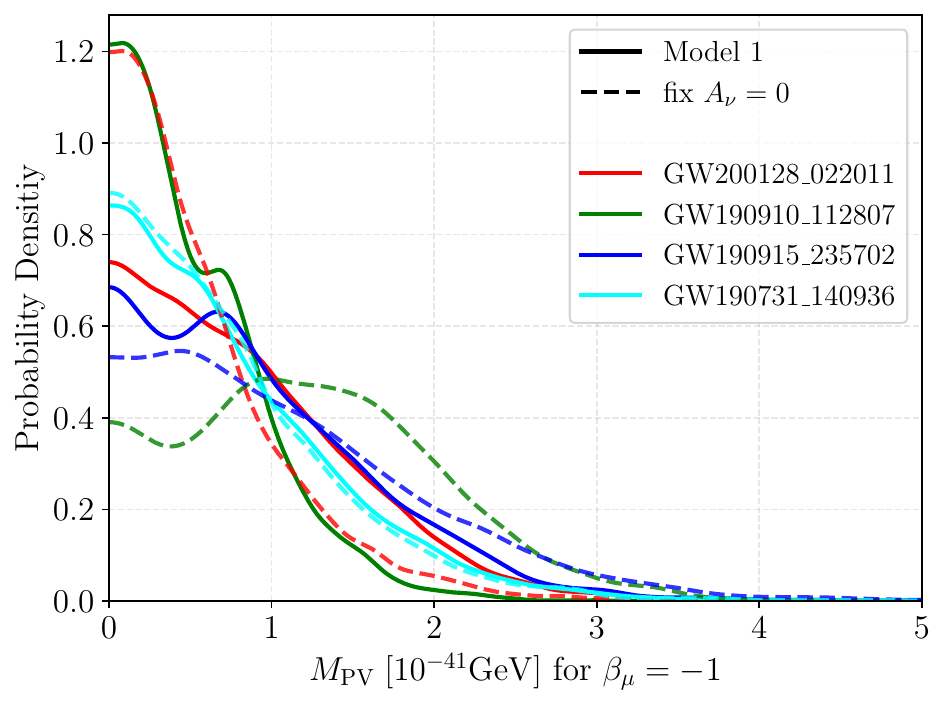}
  \includegraphics[width=0.235\textwidth]{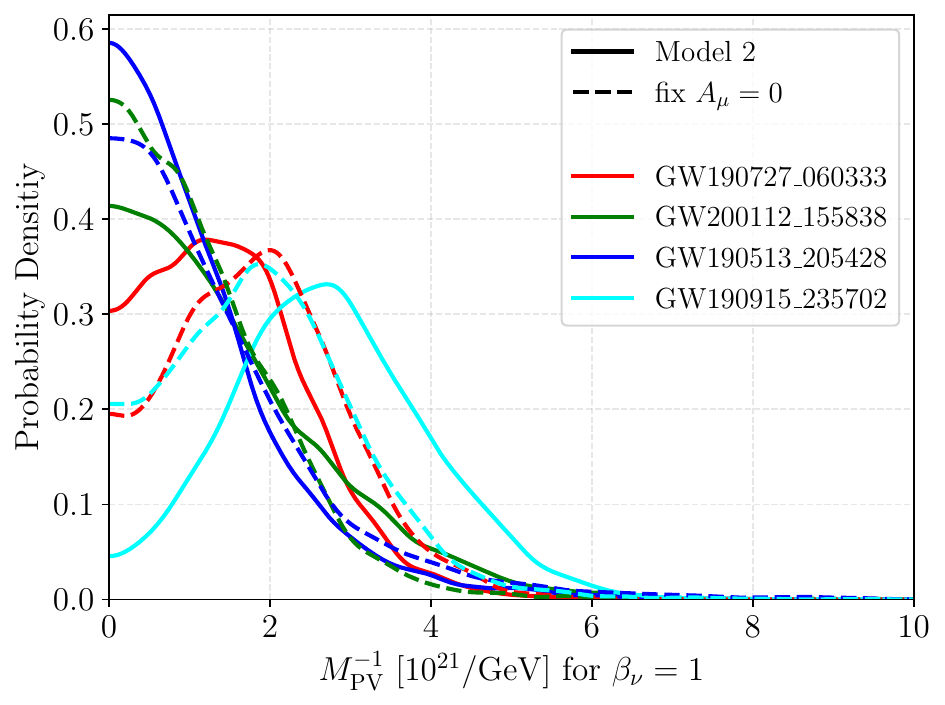}
  \includegraphics[width=0.235\textwidth]{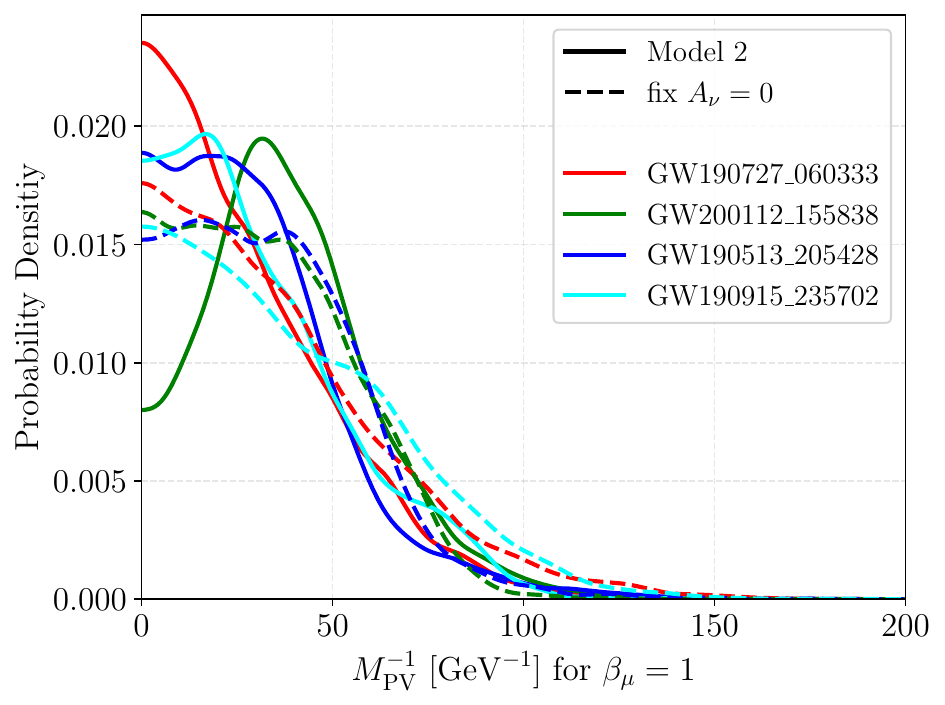}
  \includegraphics[width=0.235\textwidth]{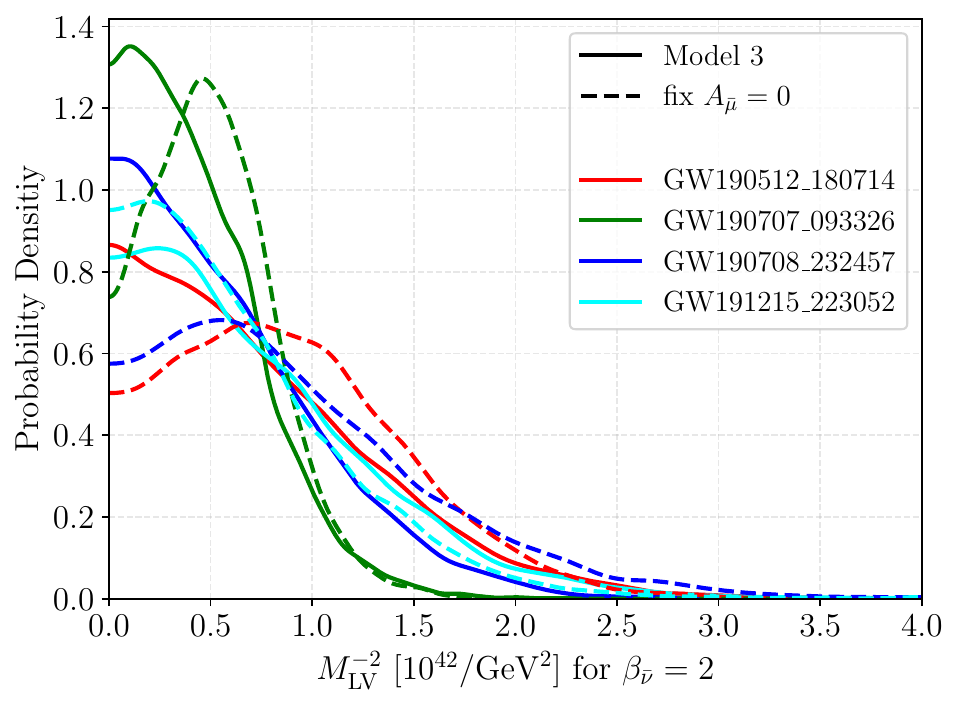}
  \includegraphics[width=0.235\textwidth]{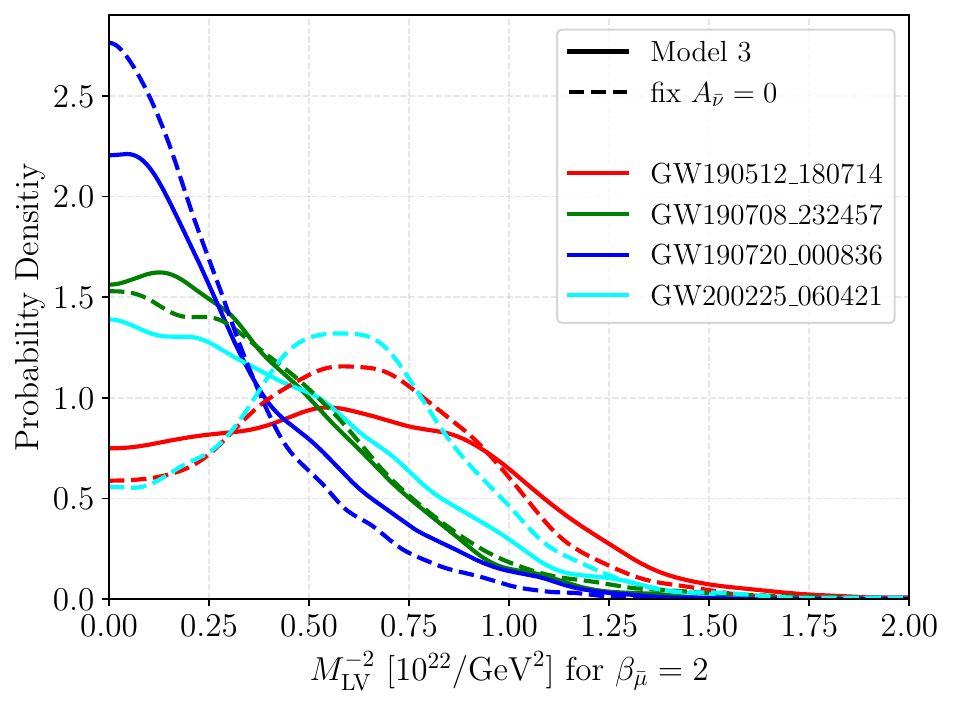}
  \includegraphics[width=0.235\textwidth]{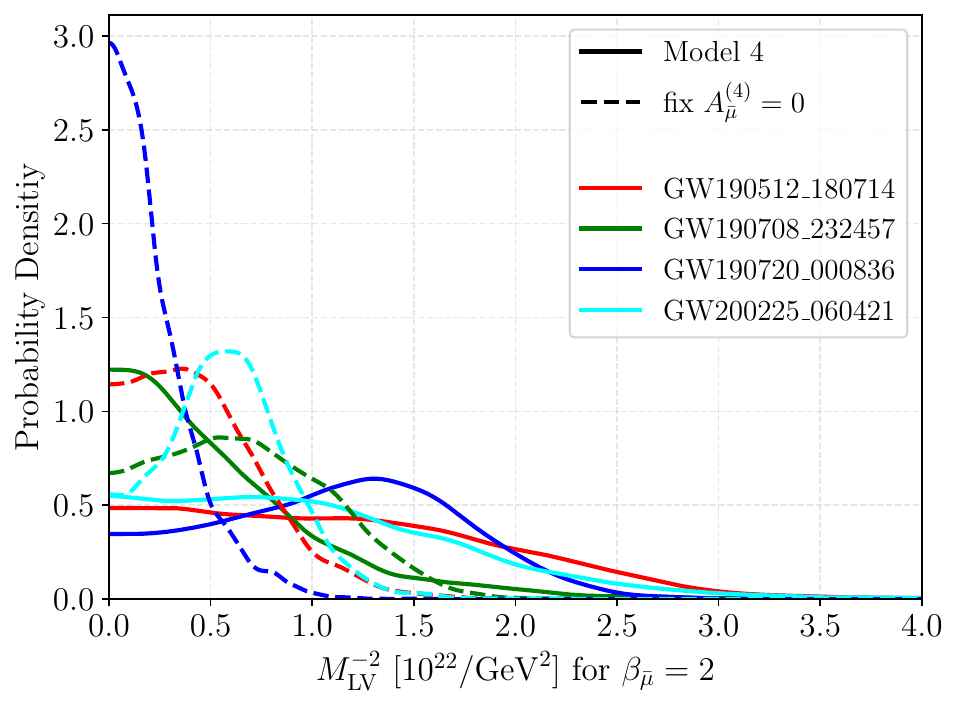}
  \includegraphics[width=0.235\textwidth]{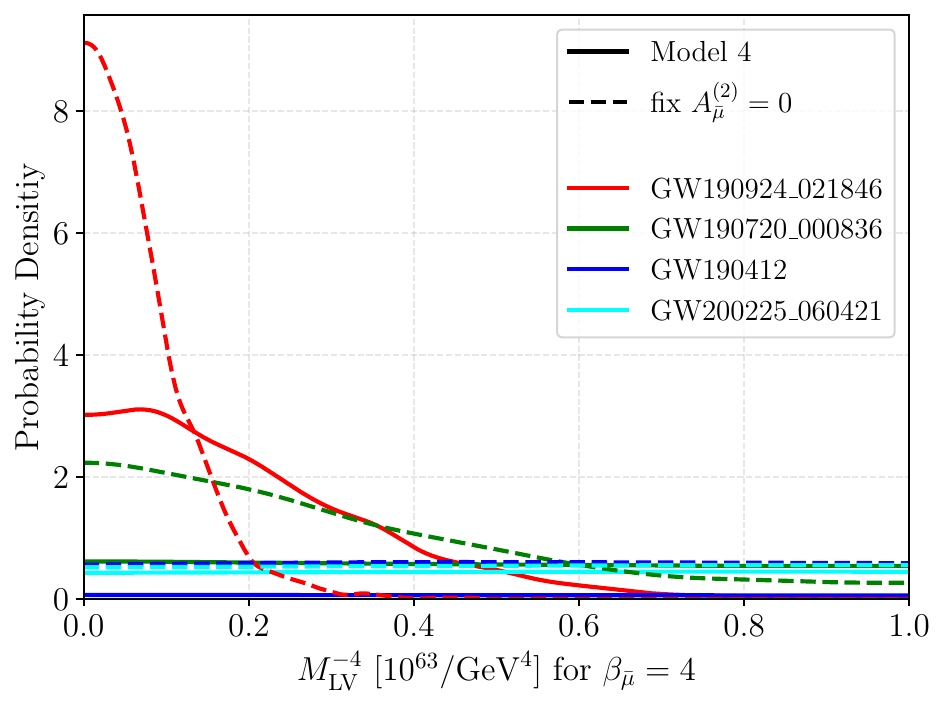}
  \caption{The posterior probability distributions of $M_{\mathrm{PV}}^{-\beta_\nu}, M_{\mathrm{PV}}^{-\beta_\mu}$, $M_{\mathrm{LV}}^{-\beta_{\bar{\nu}}}$, and $M_{\mathrm{LV}}^{-\beta_{\bar{\mu}}}$  obtained through multi-parameter analysis (solid line) and single-parameter analysis (dashed line) of the events selected from \cite{Zhu:2023rrx}, with different events represented by different colors.
  }
  \label{fig:fig}
\end{figure}

The resulting posterior distributions of $M_{\mathrm{PV}}^{-\beta_\mu}$  and $M_{\mathrm{LV}}^{-\beta_{\bar{\nu}}}, M_{\mathrm{LV}}^{-\beta_{\bar{\mu}}}$ are presented in our Figure \ref{fig:fig}. By integrating over the probability density distribution of these parameters, we derived the 90\% upper or lower limits for $M_{\mathrm{PV}}$ and $M_{\mathrm{LV}}$ and list them in Table \ref{tab:tab2}. 
{Within a given model, the constraints on $M_{\mathrm{PV}}$ or $M_{\mathrm{LV}}$ can vary by orders of magnitude depending on whether the dominant modification enters the amplitude (e.g., $\beta v=1$ ) or the phase (e.g., $\beta \mu=1$ ) of the waveform. This reflects the superior sensitivity of gravitational-wave observations to coherent phase deviations compared to overall amplitude calibrations, a well-known characteristic of matched-filtering analyses.}

\begin{figure*}[t]
  \centering
  \includegraphics[width=0.33\textwidth]{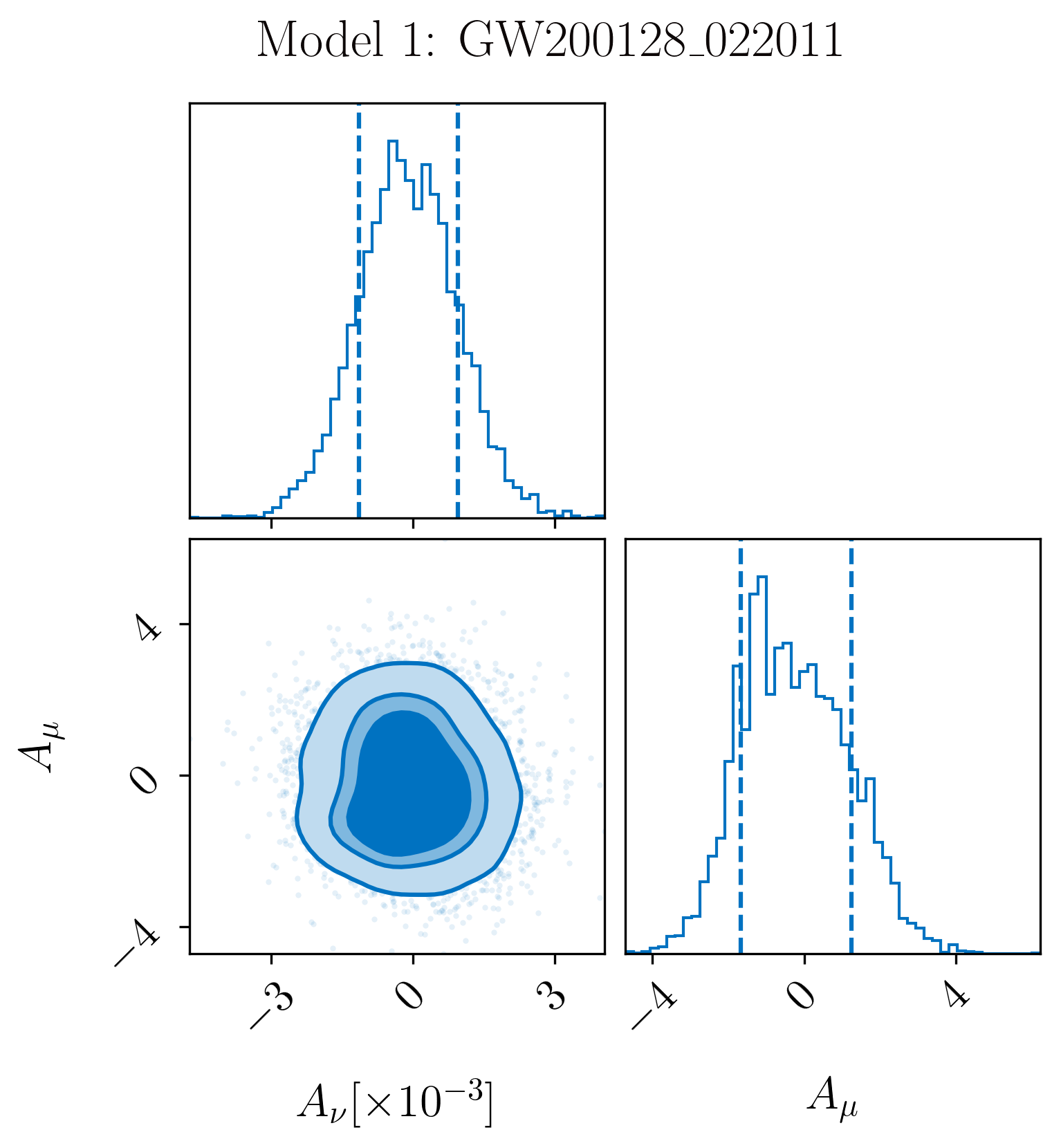}
  \includegraphics[width=0.33\textwidth]{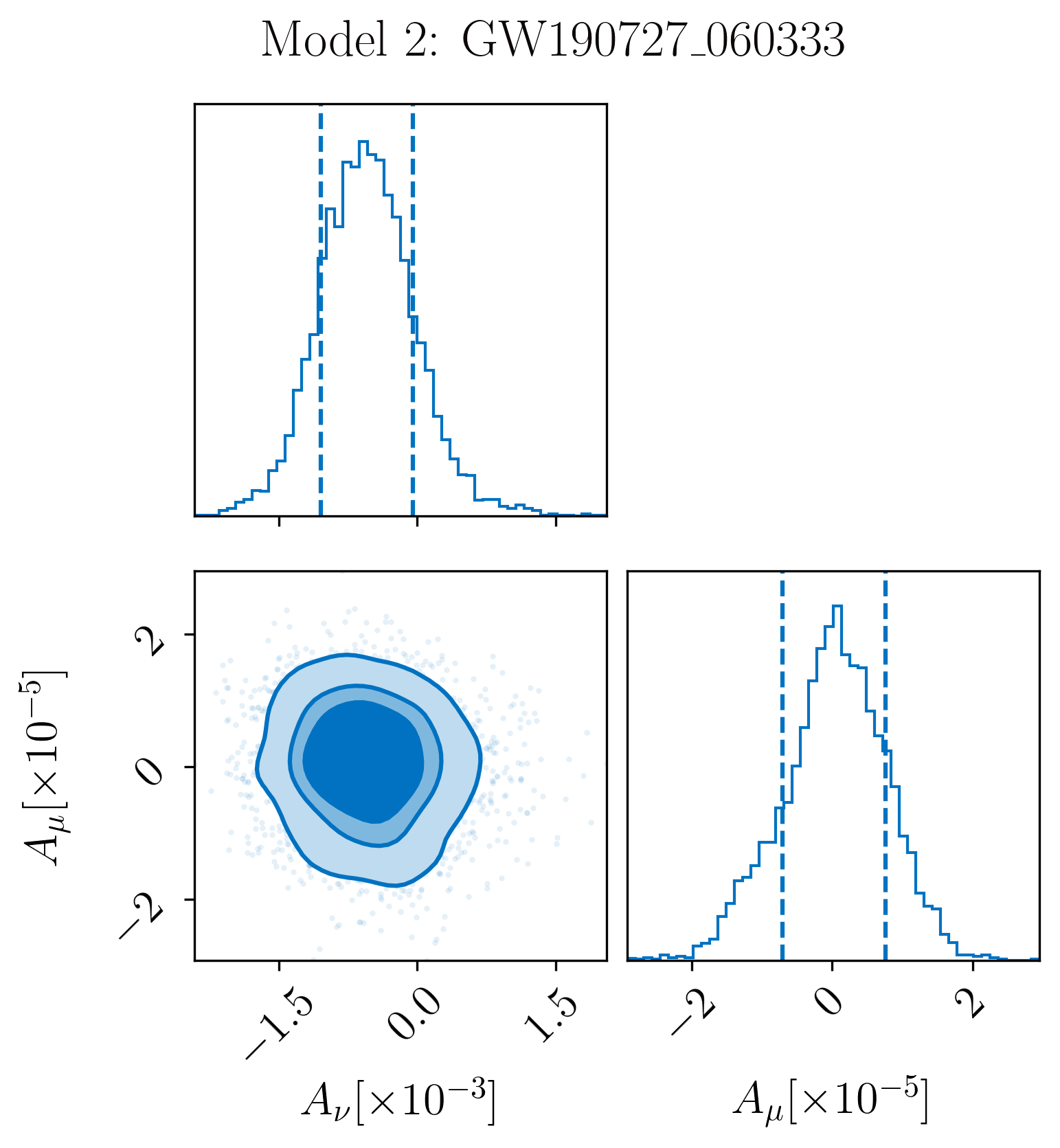}
  \includegraphics[width=0.33\textwidth]{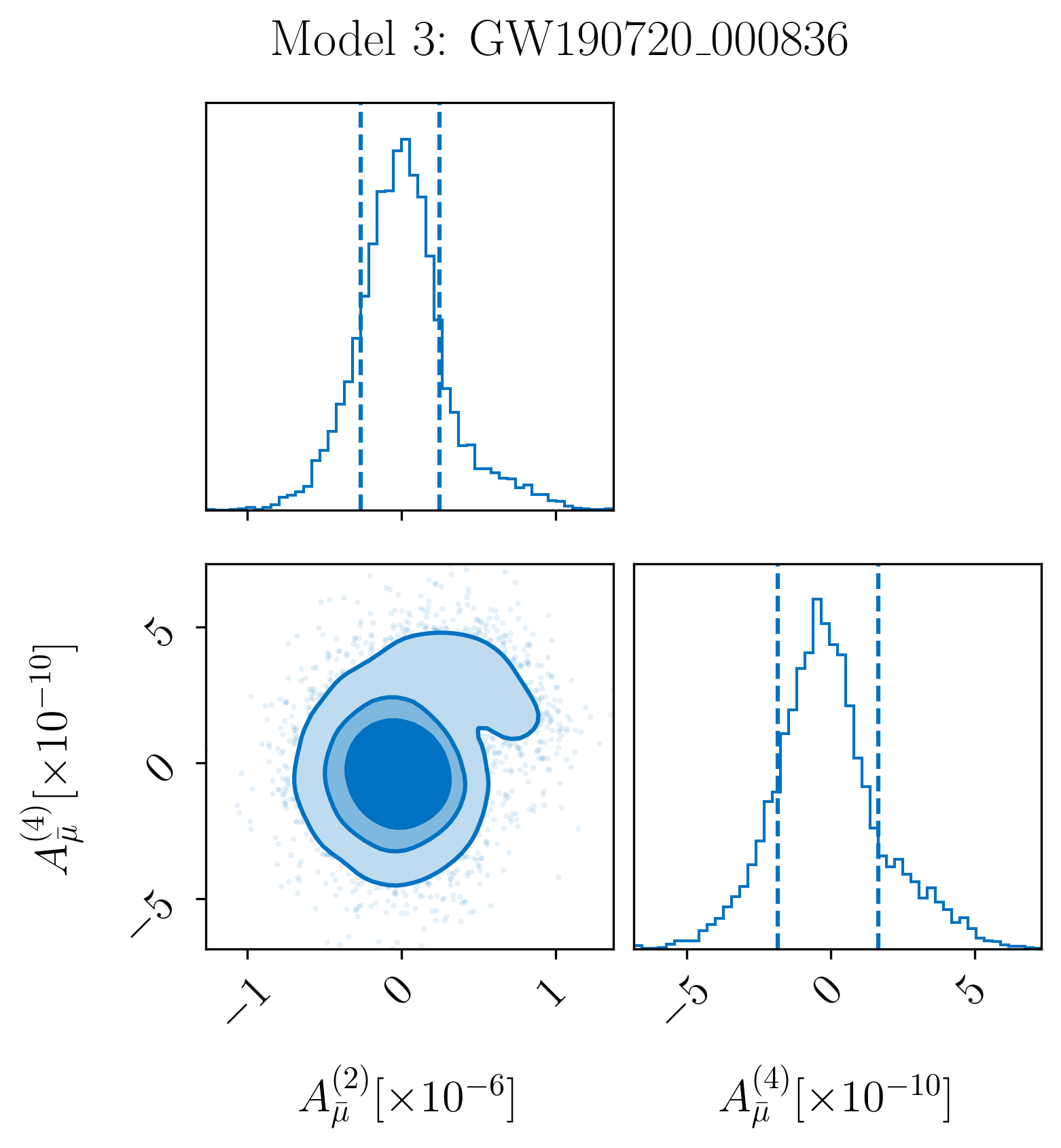}
  \includegraphics[width=0.33\textwidth]{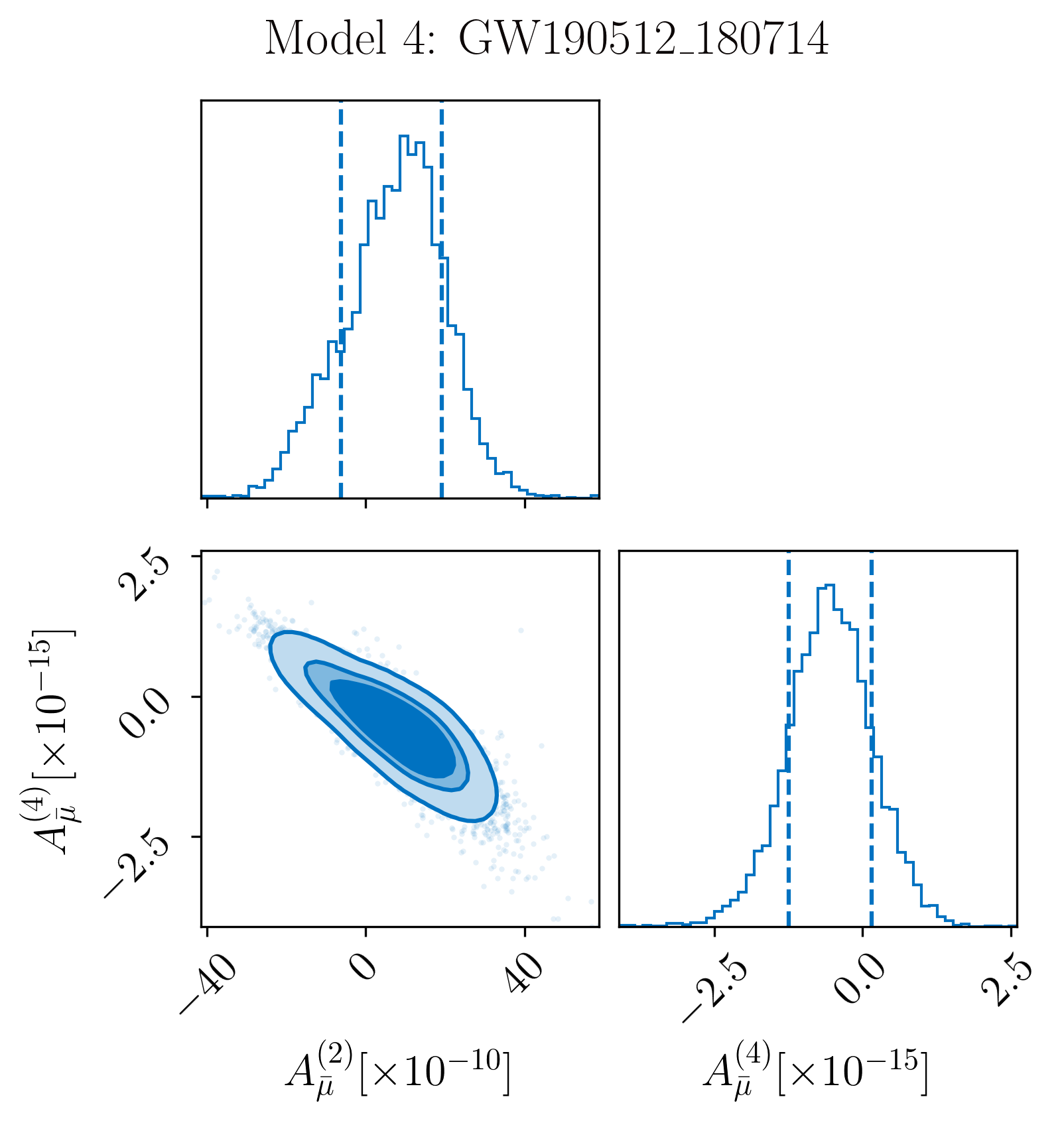}
  \caption{Corner plots of posterior distributions and correlations of parity- and Lorentz-violating amplitude parameters ($A_{\nu}$, $A_{\mu}$, $A_{\bar{\nu}}$, $A_{\bar{\mu}}$) derived from multi-parameter analysis. Each subplot represents the inference result of a dedicated event for one model. The shaded regions represent 50\%, 68\%, and 90\% credible regions, respectively.}
  \label{fig:2Dcorner}
\end{figure*}

\begin{table*}
\begin{ruledtabular}
\caption{\label{tab:tab2}Estimated $90 \%$ credible limits on parity- and Lorentz-violating mass scales: upper bounds for $M_{\mathrm{PV}}$ when $\beta_\mu=-1$ and lower bounds for $M_{\mathrm{PV}}$ or $M_{\mathrm{LV}}$ (in GeV ) for other parameter values. Constraints are derived from selected gravitational wave events across four theoretical models (Models 1-5) using both multi-parameter and single-parameter analyses.}
\begin{tabular}{ccccc}
\multirow{2}{*}{Model}   &\multirow{2}{*}{values of $\beta_\nu, \beta_\mu, \; \beta_{\bar \nu}, \beta_{\bar \mu}$}&\multirow{2}{*}{Events}& \multicolumn{2}{c}{$M_{\mathrm{PV}}$ or $M_{\mathrm{LV}}[\mathrm{GeV}]$} \\
\cline{4-5}
 &&&multi-parameter analysis &single-parameter analysis\\
 \hline
\multirow{8}{*}{Model 1}&\multirow{4}{*}{$\beta_\nu=1$} &GW$190727\_060333$  & 4.5$\times 10^{-22}$ &3.0$\times 10^{-22}$\\
&&GW$200112\_155838$  & 2.7$\times 10^{-22}$ & 4.0$\times 10^{-22}$ \\
&&GW$190513\_205428$  &3.1$\times 10^{-22}$ & 3.2$\times 10^{-22}$ \\
&&GW$200311\_115853$  & 2.8$\times 10^{-22}$& 2.6$\times 10^{-22}$ \\
\cline{2-5}
&\multirow{4}{*}{$\beta_\mu=-1$}  &GW$200128\_022011$ &  1.8$\times 10^{-41}$&1.3$\times 10^{-41}$\\
&&GW$190910\_112807$  &1.2$\times 10^{-41}$  & 2.3$\times 10^{-41}$ \\
&&GW$190915\_235702$  & 1.9$\times 10^{-41}$ & 2.3$\times 10^{-41}$ \\
&&GW$190731\_140936$  & 1.7$\times 10^{-41}$ & 1.7$\times 10^{-41}$\\
\hline
\multirow{8}{*}{Model 2}&\multirow{4}{*}{$\beta_\nu=1$}  &GW$190727\_060333$ & 3.5$\times 10^{-22}$ &3.1$\times 10^{-22}$\\
&&GW$200112\_155838$  &  3.0$\times 10^{-22}$&  4.0$\times 10^{-22}$\\
&&GW$190513\_205428$  &  3.8$\times 10^{-22}$&  3.2$\times 10^{-22}$\\
&&GW$190915\_235702$  &  2.3$\times 10^{-22}$&  2.9$\times 10^{-22}$\\
\cline{2-5}
&\multirow{4}{*}{$\beta_\mu=1$} &GW$190727\_060333$  &1.6$\times 10^{-2}$  &1.2$\times 10^{-2}$\\
&&GW$200112\_155838$  &  1.5$\times 10^{-2}$& 1.6$\times 10^{-2}$ \\
&&GW$190513\_205428$  & 1.7$\times 10^{-2}$ & 1.6$\times 10^{-2}$\\
&&GW$190915\_235702$  &1.5$\times 10^{-2}$ & 1.2$\times 10^{-2}$ \\
\hline
\multirow{8}{*}{Model 3}&\multirow{4}{*}{$\beta_{\bar{\nu}}=2$}  &GW$190512\_180714$ & 8.0$\times 10^{-22}$ &7.6$\times 10^{-22}$\\
&&GW$190707\_093326$  &  1.1$\times 10^{-21}$& 1.0$\times 10^{-21}$ \\
&&GW$190708\_232457$  &  8.9$\times 10^{-22}$&7.4$\times 10^{-22}$  \\
&&GW$191215\_223052$  & 8.1$\times 10^{-22}$ &8.5$\times 10^{-22}$ \\
\cline{2-5}
&\multirow{4}{*}{$\beta_{\bar{\mu}}=2$}  &GW$190512\_180714$ & 9.5$\times 10^{-12}$ &9.9$\times 10^{-12}$\\
&&GW$190708\_232457$  &  1.1$\times 10^{-11}$& 1.1$\times 10^{-11}$ \\
&&GW$190720\_000836$  &  1.2$\times 10^{-11}$& 1.3$\times 10^{-11}$ \\
&&GW$200225\_060421$  & 1.1$\times 10^{-11}$& 1.0$\times 10^{-11}$ \\
\hline
\multirow{8}{*}{Model  4}&\multirow{4}{*}{$\beta_{\bar{\mu},1}=2$} &GW$190512\_180714$  &6.6$\times 10^{-12}$&{1.1$\times 10^{-11}$}\\
&&GW$190708\_232457$  & 8.9$\times 10^{-12}$ & {9.0$\times 10^{-12}$} \\
&&GW$190720\_000836$  & {7.3$\times 10^{-12}$} & {1.4$\times 10^{-11}$} \\
&&GW$200225\_060421$  & 7.0$\times 10^{-12}$ &{1.0$\times 10^{-11}$} \\
\cline{2-5}
&\multirow{4}{*}{$\beta_{\bar{\mu},2}=4$} &GW$190924\_021846$  &2.2$\times 10^{-16}$ &{2.8$\times 10^{-16}$}\\
&&GW$190720\_000836$  &{1.3$\times 10^{-16}$}&{1.9$\times 10^{-16}$}  \\
&&GW$190412$  & 1.0$\times 10^{-16}$ &{1.3$\times 10^{-16}$} \\
&&GW$200225\_060421$  &1.1$\times 10^{-16}$&{1.3$\times 10^{-16}$}  \\
\hline
\multirow{5}{*}{Model  5}&\multirow{1}{*}{$\beta_\nu=1$} &GW$250114\_082203$  &7.7$\times 10^{-22}$ &{2.0$\times 10^{-21}$}\\
\cline{2-5}
&\multirow{1}{*}{$\beta_\mu=1$} 
&GW$250114\_082203$  &6.5$\times 10^{-3}$ &{1.6$\times 10^{-2}$}\\
\cline{2-5}
&\multirow{1}{*}{$\beta_{\bar{\nu}}=2$} 
&GW$250114\_082203$  &4.7$\times 10^{-22}$ &{8.0$\times 10^{-22}$}\\
\cline{2-5}
&\multirow{1}{*}{$\beta_{\bar{\mu}}=2$} 
&GW$250114\_082203$  &3.3$\times 10^{-12}$ &{5.1$\times 10^{-12}$}
\end{tabular}
\end{ruledtabular}
\end{table*}

As shown in Figure \ref{fig:fig} and Table \ref{tab:tab2}, we find that for Model 1-3, the posterior distributions obtained from multi-parameter and single-parameter analyses are roughly consistent with each other. The 90\% lower (upper) limits of $M_{\mathrm{PV}}$ and $M_{\rm LV}$ for different values of $\beta_\nu, \; \beta_{\bar \nu}, \beta_{\mu}, \beta_{\bar \mu}$ in models 1-3 are of the same order of magnitude in the two analyses. These constraints are in agreement with the results reported in ref.~\cite{Zhu:2023rrx}.

At first sight, the above results seem in contradiction with the phenomena that appeared in \cite{LIGOScientific:2016lio}: the multi-parameter analyses obtained much weaker constraints than the single-parameter analyses. As pointed out in ref.~\cite{LIGOScientific:2016lio}, significant correlations exist among early-inspiral derivation parameters. Due to the correlations, a shift in one testing parameter can always be compensated by an opposite change in another parameter while still yielding the same overall GW phase. Consequently, this leads to more conservative conclusions (weaker constraints) compared to those from single-parameter analyses. For models 1-3 in our work, the two non-GR parameters have relatively independent influences on the GW waveform (unlike ref.~\cite{LIGOScientific:2016lio} in which all the post-Newtonian coefficients describe the phase evolution), and we can check for the degeneracies between the two parameters by examining their joint posterior distributions. As shown in Figure \ref{fig:2Dcorner}, the two non-GR parameters do not correlate with each other in the inferences for models 1-3. On the other hand, $A_{\bar{\mu}}^{(2)}$ and $A_{\bar{\mu}}^{(4)}$ in model 4 correlated with each other in the corresponding inference, as both of them describe phase modifications. As shown in Table \ref{tab:tab2}, the derived constraints of $M_{\rm LV}$ from the multi-parameter analysis are a little bit weaker than those from the single-parameter analysis in this model, although they are still comparable. The degree of discrepancies can be affected by multiple factors, e.g., the available frequencies in the data, the noise abstractions, and the statistical fluctuation in the nested sampling process. For all the events we considered, the multi-parameter analyses still yield lower bounds of $M_{\rm LV}$ in the same order of magnitude compared with the single-parameter analyses.

For certain events in our analysis, we also observe that the non-GR parameters $A_\nu$ and $A_\mu$ are correlated with some of the GR parameters, leading to additional uncertainties on the constraints. For example, $A_\nu$ is anti-correlated with luminosity distance ($d_L$) in the inferred results of GW190513\_205428. This could be due to both parameters affecting the magnitude of the waveform (though with different frequency dependence), and a larger $A_\nu$ or a larger $d_L$ can both decrease the magnitude. $A_\nu$ correlates with $\theta_{JN}$ in some other events as well, as $\theta_{JN}$ and $d_L$ themselves are correlating with each other. Similar phenomena are also reported and discussed in ref.~\cite{Ng:2023jjt}. In addition, there are also correlations between $A_\mu$ and $\chi_{\rm eff}$, as both of them are related to the phase evolution of GW. This correlation is more complex and can be diverse in the parameter estimation (PE) of different events. Due to the complexity of the PE process (which is affected by noise abstraction, the nested sampling algorithms, available data that enters LVK’s sensitive band, etc.), we are unable to clarify and explain all the correlations that appear in the results of all events.

\section{Conculsion and Discussions}
\renewcommand{\theequation}{6.\arabic{equation}} \setcounter{equation}{0}

The increasing number of GW detections from compact binary coalescences by the LVK Collaboration enables unprecedented tests of gravity in the strong-field regime, with current results showing excellent agreement with GR. Anticipated advances with next-generation GW observatories will further improve sensitivity to minute deviations from GR, such as parity and Lorentz violations predicted by various modified gravity theories. Most existing analyses constrain potential deviations using a single-parameter approach, wherein all but one deformation parameter is fixed to its GR values. In this work, we investigate the differences in constraining parity and Lorentz-violating effects during gravitational-wave propagation between single- and multi-parameter approaches, within the parametric framework of \cite{Zhao:2019xmm, Zhu:2023rrx}. For this purpose, we construct several specific multi-parameter GW waveform models incorporating parity- and Lorentz-violating effects, motivated by certain specific Lorentz and parity-violating gravities. We then perform a comprehensive series of full Bayesian parameter estimation analyses for these multi-parameter waveforms, comparing the resulting constraints to those from conventional single-parameter studies. We find that when $A_\mu$ and $A_\nu$ are included in the multi-parameter approach (model 1-3), their constraints are comparable to those obtained from the single-parameter approach. This is because, within our analytical framework, these two parameters represent distinct corrections to the phase and amplitude of the waveform, respectively, and thus exhibit no significant degeneracy in the parameter estimation. In contrast, for model 4, where both non-GR parameters are associated with the same physical effect, the multi-parameter constraints become weaker than their single-parameter counterparts. Our study provides insights for assessing the robustness of single-parameter constraints within this framework. Future works could develop methods similar to those in ref.~\cite{Perkins:2022fhr} to mitigate the impact of covariances on the multi-parameter analyses where parameters describe the same effect.

\section*{Acknowledgements}

This work is supported by the National Key Research and Development Program of China under Grant No.~2020YFC2201503, the National Natural Science Foundation of China under Grants No.~12542053, No.~12275238 and No.~11675143, the Zhejiang Provincial Natural Science Foundation of China under Grants No.~LR21A050001 and No.~LY20A050002, and the Fundamental Research Funds for the Provincial Universities of Zhejiang in China under Grant No.~RF-A2019015. Y-Z.W. is supported by the National Natural Science Foundation of China (Grant No.~12203101).

This research has made use of data or software obtained from the Gravitational Wave Open Science Center (gwosc.org), a service of the LIGO Scientific Collaboration, the Virgo Collaboration, and KAGRA. This material is based upon work supported by NSF's LIGO Laboratory which is a major facility fully funded by the National Science Foundation, as well as the Science and Technology Facilities Council (STFC) of the United Kingdom, the Max-Planck-Society (MPS), and the State of Niedersachsen/Germany for support of the construction of Advanced LIGO and construction and operation of the GEO600 detector. Additional support for Advanced LIGO was provided by the Australian Research Council. Virgo is funded, through the European Gravitational Observatory (EGO), by the French Centre National de Recherche Scientifique (CNRS), the Italian Istituto Nazionale di Fisica Nucleare (INFN) and the Dutch Nikhef, with contributions by institutions from Belgium, Germany, Greece, Hungary, Ireland, Japan, Monaco, Poland, Portugal, Spain. KAGRA is supported by the Ministry of Education, Culture, Sports, Science and Technology (MEXT), Japan Society for the Promotion of Science (JSPS) in Japan; National Research Foundation (NRF) and Ministry of Science and ICT (MSIT) in Korea; Academia Sinica (AS) and National Science and Technology Council (NSTC) in Taiwan.

The data analyses and results visualization in this work made use of \texttt{BILBY} \cite{Romero-Shaw:2020owr, Ashton:2018jfp}, \texttt{NESSAI} \cite{2021PhRvD.103j3006W}, \texttt{LALSuite} \cite{LALSuite}, \texttt{Numpy} \cite{Harris:2020xlr, vanderWalt:2011bqk}, \texttt{Scipy} \cite{Virtanen:2019joe}, and \texttt{matplotlib} \cite{Hunter:2007ouj}.

\clearpage

\appendix

\section{ The prior distributions}

 In this appendix, we present the priors of the parameter we analyzed in our analysis. The priors of the chirp mass ${\cal M}$, luminosity distance $d_L$, the mass ratio $q$, and the non-GR parameters in the analysis of each GW event are summarized in Table \ref{tab:tab3}.

\begin{table*}
\tiny
\caption{\label{tab:tab3}Prior distribution intervals for a subset of parameters. m and s denote multi-parameter and single-parameter analysis methods, respectively.}
\begin{ruledtabular}
\begin{tabular}{cccclllll}
Model  & values of $\beta_\nu, \beta_\mu, \beta_{\bar{\nu}}, \beta_{\bar{\mu}}$ &Method &Events &$q$&$\mathcal{M}$(M$_\odot$)&$d_L$ (Mpc)& \multicolumn{2}{c}{Non-GR Parameters}\\
 \hline															
 &&&&&&&$A_\nu\times10^{-2}$&$A_\mu\times10^{1}$\\	
 \cline{8-9}			
 \multirow{8}{*}{Model 1}	&	\multirow{8}{*}{$\beta_\nu=1$} 	&	\multirow{4}{*}{m}  	&	GW$190727\_060333$	&	[	0.03	,	1	]	&	[	26.7	,	58.7	]	&	[	218.3	,	10000	]	&	[	-5	,	5	]	&	[	-2	,	2	]	\\
	&		&		&	GW$200112\_155838$	&	[	0.03	,	1	]	&	[	25.9	,	43.1	]	&	[	220.2	,	2278.6	]	&	[	-5	,	5	]	&	[	-2	,	2	]	\\
	&		&		&	GW$190513\_205428$	&	[	0.03	,	1	]	&	[	15.3	,	47.9	]	&	[	156.6	,	5000	]	&	[	-5	,	5	]	&	[	-2	,	2	]	\\
	&		&		&	GW$200311\_115853$	&	[	0.125	,	1	]	&	[	22.5	,	41.3	]	&	[	373.3	,	1927.4	]	&	[	-5	,	5	]	&	[	-2	,	2	]	\\
\cline{3-9}					
	&		&	\multirow{4}{*}{s} 	&	GW$190727\_060333$	&	[	0.03	,	1	]	&	[	26.7	,	58.7	]	&	[	218.3	,	10000	]	&	[	-5	,	5	]	&	[	0	,	0	]	\\
	&		&		&	GW$200112\_155838$	&	[	0.03	,	1	]	&	[	25.9	,	43.1	]	&	[	220.2	,	2278.6	]	&	[	-5	,	5	]	&	[	0	,	0	]	\\
	&		&		&	GW$190513\_205428$	&	[	0.03	,	1	]	&	[	15.3	,	47.9	]	&	[	156.6	,	5000	]	&	[	-5	,	5	]	&	[	0	,	0	]	\\
	&		&		&	GW$200311\_115853$	&	[	0.125	,	1	]	&	[	22.5	,	41.3	]	&	[	373.3	,	1927.4	]	&	[	-5	,	5	]	&	[	0	,	0	]	\\
 \hline					
 &&&&&&& $A_\nu\times10^{-2}$&$A_\mu\times10^{1}$\\				
 \cline{8-9}												
 \multirow{8}{*}{Model 1}	&	\multirow{8}{*}{$\beta_\mu=-1$} 	&	\multirow{4}{*}{m}  	&	GW$200128\_022011$	&	[	0.05	,	1	]	&	[	29.2	,	64.4	]	&	[	468	,	8485.2	]	&	[	-5	,	5	]	&	[	-2	,	2	]	\\
	&		&		&	GW$190910\_112807$	&	[	0.125	,	1	]	&	[	20	,	80	]	&	[	20	,	5000	]	&	[	-5	,	5	]	&	[	-2	,	2	]	\\
	&		&		&	GW$190915\_235702$	&	[	0.03	,	1	]	&	[	20	,	41.3	]	&	[	42.4	,	3500	]	&	[	-5	,	5	]	&	[	-2	,	2	]	\\
	&		&		&	GW$190731\_140936$	&	[	0.125	,	1	]	&	[	25	,	75	]	&	[	100	,	10000	]	&	[	-5	,	5	]	&	[	-2	,	2	]	\\
\cline{3-9}													
	&		&	\multirow{4}{*}{s} 	&	GW$200128\_022011$	&	[	0.05	,	1	]	&	[	29.2	,	64.4	]	&	[	468	,	8485.2	]	&	[	0	,	0	]	&	[	-2	,	2	]	\\
	&		&		&	GW$190910\_112807$	&	[	0.125	,	1	]	&	[	20	,	80	]	&	[	20	,	5000	]	&	[	0	,	0	]	&	[	-2	,	2	]	\\
	&		&		&	GW$190915\_235702$	&	[	0.03	,	1	]	&	[	20	,	41.3	]	&	[	42.4	,	3500	]	&	[	0	,	0	]	&	[	-2	,	2	]	\\
	&		&		&	GW$190731\_140936$	&	[	0.125	,	1	]	&	[	25	,	75	]	&	[	100	,	10000	]	&	[	0	,	0	]	&	[	-2	,	2	]	\\
 \hline						
&&&&&&&$A_\nu\times10^{-2}$&$A_\mu\times10^{-2}$\\				
 \cline{8-9}													
 \multirow{8}{*}{Model 2}	&	\multirow{8}{*}{$\beta_\nu=1$} 	&	\multirow{4}{*}{m}  	&	GW$190727\_060333$	&	[	0.125	,	1	]	&	[	25	,	100	]	&	[	100	,	5000	]	&	[	-5	,	5	]	&	[	-0.08	,	0.08	]	\\
	&		&		&	GW$200112\_155838$	&	[	0.03	,	1	]	&	[	25.9	,	43.1	]	&	[	220.2	,	2278.6	]	&	[	-5	,	5	]	&	[	-5	,	5	]	\\
	&		&		&	GW$190513\_205428$	&	[	0.03	,	1	]	&	[	15.3	,	47.9	]	&	[	156.6	,	5000	]	&	[	-5	,	5	]	&	[	-5	,	5	]	\\
	&		&		&	GW$190915\_235702$	&	[	0.03	,	1	]	&	[	20	,	41.3	]	&	[	42.4	,	3500	]	&	[	-5	,	5	]	&	[	-5	,	5	]	\\
\cline{3-9}																		
	&		&	\multirow{4}{*}{s}  	&	GW$190727\_060333$	&	[	0.125	,	1	]	&	[	25	,	100	]	&	[	100	,	5000	]	&	[	-5	,	5	]	&	[	0	,	0	]	\\
	&		&		&	GW$200112\_155838$	&	[	0.03	,	1	]	&	[	25.9	,	43.1	]	&	[	220.2	,	2278.6	]	&	[	-5	,	5	]	&	[	0	,	0	]	\\
	&		&		&	GW$190513\_205428$	&	[	0.03	,	1	]	&	[	15.3	,	47.9	]	&	[	156.6	,	5000	]	&	[	-5	,	5	]	&	[	0	,	0	]	\\
	&		&		&	GW$190915\_235702$	&	[	0.03	,	1	]	&	[	20	,	41.3	]	&	[	42.4	,	3500	]	&	[	-5	,	5	]	&	[	0	,	0	]	\\
 \hline														
&&&&&&&$A_\nu\times10^{-2}$&$A_\mu\times10^{-2}$\\			
 \cline{8-9}													
 \multirow{8}{*}{Model 2}	&	\multirow{8}{*}{$\beta_\mu=1$} 	&	\multirow{4}{*}{m}  	&	GW$190727\_060333$	&	[	0.125	,	1	]	&	[	25	,	100	]	&	[	100	,	5000	]	&	[	-5	,	5	]	&	[	-0.08	,	0.08	]	\\
	&		&		&	GW$200112\_155838$	&	[	0.03	,	1	]	&	[	25.9	,	43.1	]	&	[	220.2	,	2278.6	]	&	[	-5	,	5	]	&	[	-5	,	5	]	\\
	&		&		&	GW$190513\_205428$	&	[	0.03	,	1	]	&	[	15.3	,	47.9	]	&	[	156.6	,	5000	]	&	[	-5	,	5	]	&	[	-5	,	5	]	\\
	&		&		&	GW$190915\_235702$	&	[	0.03	,	1	]	&	[	20	,	41.3	]	&	[	42.4	,	3500	]	&	[	-5	,	5	]	&	[	-5	,	5	]	\\
\cline{3-9}																																					
	&		&	\multirow{4}{*}{s}  	&	GW$190727\_060333$	&	[	0.03	,	1	]	&	[	26.7	,	58.7	]	&	[	218.3	,	10000	]	&	[	-0.005	,	0.005	]	&	[	0	,	0	]	\\
	&		&		&	GW$200112\_155838$	&	[	0.03	,	1	]	&	[	25.9	,	43.1	]	&	[	220.2	,	2278.6	]	&	[	-5	,	5	]	&	[	0	,	0	]	\\
	&		&		&	GW$190513\_205428$	&	[	0.03	,	1	]	&	[	15.3	,	47.9	]	&	[	156.6	,	5000	]	&	[	-5	,	5	]	&	[	0	,	0	]	\\
	&		&		&	GW$190915\_235702$	&	[	0.03	,	1	]	&	[	20	,	41.3	]	&	[	42.4	,	3500	]	&	[	-5	,	5	]	&	[	0	,	0	]	\\
 \hline																																					
&&&&&&&$A_{\bar{\nu}}\times10^{-3}$&$A_{\bar{\mu}}\times10^{-7}$\\																																					
 \cline{8-9}																																					
 \multirow{8}{*}{Model 3}	&	\multirow{8}{*}{$\beta_{\bar{\nu}}=2$} 	&	\multirow{4}{*}{m}  	&	GW$190512\_180714$	&	[	0.03	,	1	]	&	[	10.3	,	28.3	]	&	[	46.4	,	4000	]	&	[	-2	,	2	]	&	[	-5	,	5	]	\\
	&		&		&	GW$190707\_093326$	&	[	0.03	,	1	]	&	[	9	,	11	]	&	[	35.7	,	1700	]	&	[	-2	,	2	]	&	[	-5	,	5	]	\\
	&		&		&	GW$190708\_232457$	&	[	0.05	,	1	]	&	[	14.5	,	16.3	]	&	[	157.2	,	1637.2	]	&	[	-2	,	2	]	&	[	-5	,	5	]	\\
	&		&		&	GW$191215\_223052$	&	[	0.05	,	1	]	&	[	20.3	,	30.6	]	&	[	425	,	4299	]	&	[	-2	,	2	]	&	[	-5	,	5	]	\\
\cline{3-9}																																					
	&		&	\multirow{4}{*}{s} 	&	GW$190512\_180714$	&	[	0.03	,	1	]	&	[	10.3	,	28.3	]	&	[	46.4	,	4000	]	&	[	-2	,	2	]	&	[	0	,	0	]	\\
	&		&		&	GW$190707\_093326$	&	[	0.03	,	1	]	&	[	9	,	11	]	&	[	35.7	,	1700	]	&	[	-2	,	2	]	&	[	0	,	0	]	\\
	&		&		&	GW$190708\_232457$	&	[	0.05	,	1	]	&	[	14.5	,	16.3	]	&	[	157.2	,	1637.1	]	&	[	-2	,	2	]	&	[	0	,	0	]	\\
	&		&		&	GW$191215\_223052$	&	[	0.05	,	1	]	&	[	20.3	,	30.6	]	&	[	425	,	4299	]	&	[	-2	,	2	]	&	[	0	,	0	]	\\
 \hline																																					
&&&&&&&$A_{\bar{\nu}}\times10^{-3}$&$A_{\bar{\mu}}\times10^{-7}$\\																																					
 \cline{8-9}																																					
 \multirow{8}{*}{Model 3}	&	\multirow{8}{*}{$\beta_{\bar{\mu}}=2$} 	&	\multirow{4}{*}{m}  	&	GW$190512\_180714$	&	[	0.03	,	1	]	&	[	10.3	,	28.3	]	&	[	46.4	,	4000	]	&	[	-2	,	2	]	&	[	-5	,	5	]	\\
	&		&		&	GW$190708\_232457$	&	[	0.05	,	1	]	&	[	14.5	,	16.3	]	&	[	157.2	,	1637.2	]	&	[	-2	,	2	]	&	[	-5	,	5	]	\\
	&		&		&	GW$190720\_000836$	&	[	0.05	,	1	]	&	[	9.9	,	10.9	]	&	[	200	,	2000	]	&	[	-2	,	2	]	&	[	-5	,	5	]	\\
	&		&		&	GW$200225\_060421$	&	[	0.125	,	1	]	&	[	12.9	,	20.5	]	&	[	148.2	,	2471.3	]	&	[	-2	,	2	]	&	[	-5	,	5	]	\\
\cline{3-9}																																					
	&		&	\multirow{4}{*}{s} 	&	GW$190512\_180714$	&	[	0.03	,	1	]	&	[	10.3	,	28.3	]	&	[	46.4	,	4000	]	&	[	0	,	0	]	&	[	-0.1	,	0.1	]	\\
	&		&		&	GW$190708\_232457$	&	[	0.05	,	1	]	&	[	14.5	,	16.3	]	&	[	157.2	,	1637.2	]	&	[	0	,	0	]	&	[	-5	,	5	]	\\
	&		&		&	GW$190720\_000836$	&	[	0.05	,	1	]	&	[	9.9	,	10.9	]	&	[	200	,	2000	]	&	[	0	,	0	]	&	[	-5	,	5	]	\\
	&		&		&	GW$200225\_060421$	&	[	0.05	,	1	]	&	[	12.9	,	20.5	]	&	[	148.2	,	2471.3	]	&	[	0	,	0	]	&	[	-5	,	5	]	\\
 \hline																																					
&&&&&&&$A_{\bar{\mu}}^{(2)}\times10^{-7}$&$A_{\bar{\mu}}^{(4)}\times10^{-15}$\\																																					
 \cline{8-9}																																					
 \multirow{8}{*}{Model 4}	&	\multirow{8}{*}{$\beta_{\bar{\mu}}=2$} 	&	\multirow{4}{*}{m}  	&	GW$190512\_180714$	&	[	0.05	,	1	]	&	[	16.3	,	21.5	]	&	[	467.3	,	2857.2	]	&	[	-5	,	5	]	&	[	-50	,	50	]	\\
	&		&		&	GW$190708\_232457$	&	[	0.125	,	1	]	&	[	11	,	16	]	&	[	300	,	3000	]	&	[	-5	,	5	]	&	[	-5	,	5	]	\\
	&		&		&	GW$190720\_000836$	&	[	0.05	,	1	]	&	[	9.9	,	10.9	]	&	[	200	,	2000	]	&	[	-5	,	5	]	&	[	-5	,	5	]	\\
	&		&		&	GW$200225\_060421$	&	[	0.05	,	1	]	&	[	12.9	,	20.5	]	&	[	148.2	,	2471.3	]	&	[	-5	,	5	]	&	[	-500	,	500	]	\\
\cline{3-9}																																					
	&		&	\multirow{4}{*}{s} 	&	GW$190512\_180714$	&	[	0.05	,	1	]	&	[	16.3	,	21.5	]	&	[	467.3	,	2857.2	]	&	[	-5	,	5	]	&	[	0	,	0	]	\\
	&		&		&	GW$190708\_232457$	&	[	0.05	,	1	]	&	[	14.5	,	16.3	]	&	[	157.2	,	1637.2	]	&	[	-5	,	5	]	&	[	0	,	0	]	\\
	&		&		&	GW$190720\_000836$	&	[	0.05	,	1	]	&	[	9.9	,	10.9	]	&	[	200	,	2000	]	&	[	-5	,	5	]	&	[	0	,	0	]	\\
	&		&		&	GW$200225\_060421$	&	[	0.05	,	1	]	&	[	12.9	,	20.5	]	&	[	148.2	,	2471.3	]	&	[	-5	,	5	]	&	[	0	,	0	]	\\
 \hline																																					
&&&&&&&$A_{\bar{\mu}}^{(2)}\times10^{-7}$&$A_{\bar{\mu}}^{(4)}\times10^{-15}$\\																																					
 \cline{8-9}																																					
 \multirow{8}{*}{Model 4}	&	\multirow{8}{*}{$\beta_{\bar{\mu}}=4$} 	&	\multirow{4}{*}{m}  	&	GW$190924\_021846$	&	[	0.125	,	1	]	&	[	4	,	8	]	&	[	250	,	900	]	&	[	-5	,	5	]	&	[	-5	,	5	]	\\
	&		&		&	GW$190720\_000836$	&	[	0.125	,	1	]	&	[	5	,	14	]	&	[	300	,	2000	]	&	[	-5	,	5	]	&	[	-5	,	5	]	\\
	&		&		&	GW$190412$	&	[	0.125	,	1	]	&	[	10	,	16	]	&	[	100	,	2000	]	&	[	-5	,	5	]	&	[	-50	,	50	]	\\
	&		&		&	GW$200225\_060421$	&	[	0.05	,	1	]	&	[	12.9	,	20.5	]	&	[	148.2	,	2471.3	]	&	[	-5	,	5	]	&	[	-500	,	500	]	\\
\cline{3-9}																																					
	&		&	\multirow{4}{*}{s} 	&	GW$190924\_021846$	&	[	0.125	,	1	]	&	[	4	,	8	]	&	[	250	,	900	]	&	[	0	,	0	]	&	[	-5	,	5	]	\\
	&		&		&	GW$190720\_000836$	&	[	0.125	,	1	]	&	[	5	,	14	]	&	[	300	,	2000	]	&	[	0	,	0	]	&	[	-5	,	5	]	\\
	&		&		&	GW$190412$	&	[	0.125	,	1	]	&	[	10	,	16	]	&	[	100	,	2000	]	&	[	0	,	0	]	&	[	-50	,	50	]	\\
	&		&		&	GW$200225\_060421$	&	[	0.125	,	1	]	&	[	12.9	,	20.5	]	&	[	148.2	,	2471.3	]	&	[	0	,	0	]	&	[	-500	,	500	]	\\  
\end{tabular}
\end{ruledtabular}
\end{table*}

\begin{table*}
\tiny
\caption{\label{tab:tab4}Same as Table\ref{tab:tab2} but for Model 5.}
\begin{ruledtabular}
\begin{tabular}{cccclllllll}
Model  &values of $\beta_\nu, \beta_\mu, \beta_{\bar{\nu}}, \beta_{\bar{\mu}}$  &Method &Events &$q$&$\mathcal{M}$(M$_\odot$)&$d_L$ (Mpc)& \multicolumn{2}{c}{Non-GR Parameters}\\			
 \hline	
&&&&&&&$A_{\nu}\times10^{-2}$&$A_{\mu}\times10^{-4}$&$A_{\bar{\nu}}\times10^{-3}$&$A_{\bar{\mu}}\times10^{-7}$\\	
\cline{8-11}													
 \multirow{6}{*}{Model 5}	&		&	\multirow{1}{*}{m}  	&	\multirow{6}{*}{GW$250114\_082203$ }	&	\multirow{6}{*}{[	0.17	,	1	]}	&	\multirow{6}{*}{[	30	,	32	]}	&	\multirow{6}{*}{[	10	,	10000	]}	&	[	-5	,	5	]	&	[	-8	,	8	]	&[	-2	,	2	]&[	-5	,	5	]\\
\cline{2-3}
\cline{8-11}														
&		$\beta_{\nu}=1$ &	\multirow{1}{*}{s}  	&	 	&		&		&		&	[	-5	,	5	]	&	[	0	,	0	]	&[	0	,	0	]&[	0	,	0	]\\

\cline{2-3}
\cline{8-11}														
&		$\beta_{\mu}=1$ &	\multirow{1}{*}{s}  	&	 	&		&		&		&	[	0	,	0	]	&	[	-8	,	8	]	&[	0	,	0	]&[	0	,	0	]\\

\cline{2-3}
\cline{8-11}														
&		$\beta_{\bar{\nu}}=2$ &	\multirow{1}{*}{s}  	&	 	&		&		&		&	[	0	,	0	]	&	[	0	,	0	]	&[	-2	,	2	] &[	0	,	0	]\\

\cline{2-3}
\cline{8-11}														
&		$\beta_{\bar{\mu}}=2$ &	\multirow{1}{*}{s}  	&	 	&		&		&		&	[	0	,	0	]	&	[	0	,	0	]	&[	0	,	0	]&[	-5	,	5	]\\
    
\end{tabular}
\end{ruledtabular}
\end{table*}

\section{Four-Parameter Analysis with GW250114\_082203}\label{sec:4p}

{The binary black hole signal GW250114\_082203 is the loudest event detected to date \cite{LIGOScientific:2025obp}, offering us a unique opportunity to test the violations of GR. To study whether our main conclusion holds in more complicated cases, we introduce Model 5 ($\beta_\nu=1, \beta_\mu=1, \beta_{\bar{\nu}}=2$ and $\beta_{\bar{\mu}}=2$) in which both PV and LV are included below.}

{
This case can arise from the spatial covariant gravities \cite{Gao:2019liu}. This case includes four distinct effects: (1) amplitude birefringence; (2) velocity birefringence; (3) frequency-dependent damping; and (4) nonlinear dispersion. The modified waveforms for this case can be written as
\begin{eqnarray}
 \tilde h_A(f) = \tilde h_A^{\rm GR}(f) e^{ (\rho_A \delta h_1 + \delta h_2)} e^{i (\rho_A \delta \Psi_1 + \delta \Psi_2) },\label{waveforms}
\end{eqnarray}
with the amplitude and phase corrections being given by
\begin{eqnarray}
\delta h_1 = A_{\nu} \pi f, \\
\delta \Psi_1 = A_\mu (\pi f)^2,\\
\delta h_2 = - A_{\bar \nu} (\pi f)^2, \\
\delta \Psi_2 = A_{\bar \mu} (\pi f)^3.
\end{eqnarray}
Here $A_\nu$, $A_\mu$, $A_{\bar{\nu}}$ and $A_{\bar{\mu}}$ are given by 
\begin{eqnarray}
A_\nu &=& 
- \frac{\alpha_\nu}{M_{\rm PV}} z,\\
A_\mu &=& \frac{\alpha_\mu}{M_{\mathrm{PV}}} \int_0^z \frac{1+z^{\prime}}{H_0 \sqrt{\Omega_m\left(1+z^{\prime}\right)^3+\Omega_{\Lambda}}} d z^{\prime},\\
A_{\bar \nu}  &=& \frac{2}{M^2_{\mathrm{LV}}} \Big[\alpha_{\bar \nu} - \alpha_{\bar \nu} (1+z)^2\Big],\\
A_{\bar \mu} &=& \frac{8}{3}\frac{ \alpha_{\bar \mu}}{M^2_{\mathrm{LV}}} \int_0^z \frac{(1+z^{\prime})^2}{H_0 \sqrt{\Omega_m\left(1+z^{\prime}\right)^3+\Omega_{\Lambda}}} d z^{\prime}.\nb\\
\end{eqnarray}
Note that in the above we have treated $\alpha_\nu$, $\alpha_\mu$, $\alpha_{\bar \nu}$ and $\alpha_{\bar \mu}$ as constants. }

\begin{figure}
  \centering
  \includegraphics[width=0.235\textwidth]{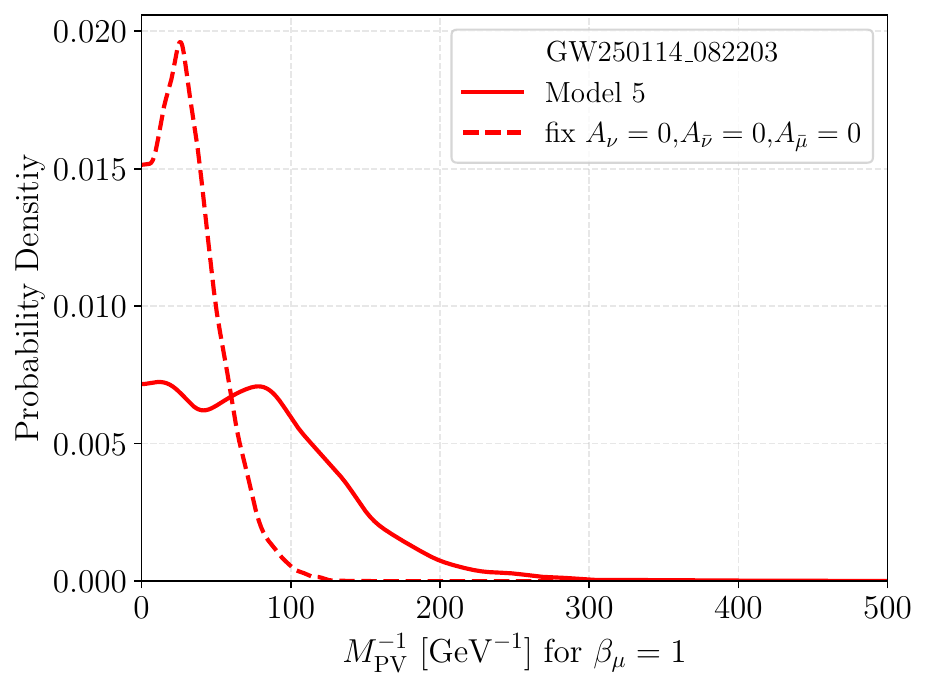}
  \includegraphics[width=0.235\textwidth]{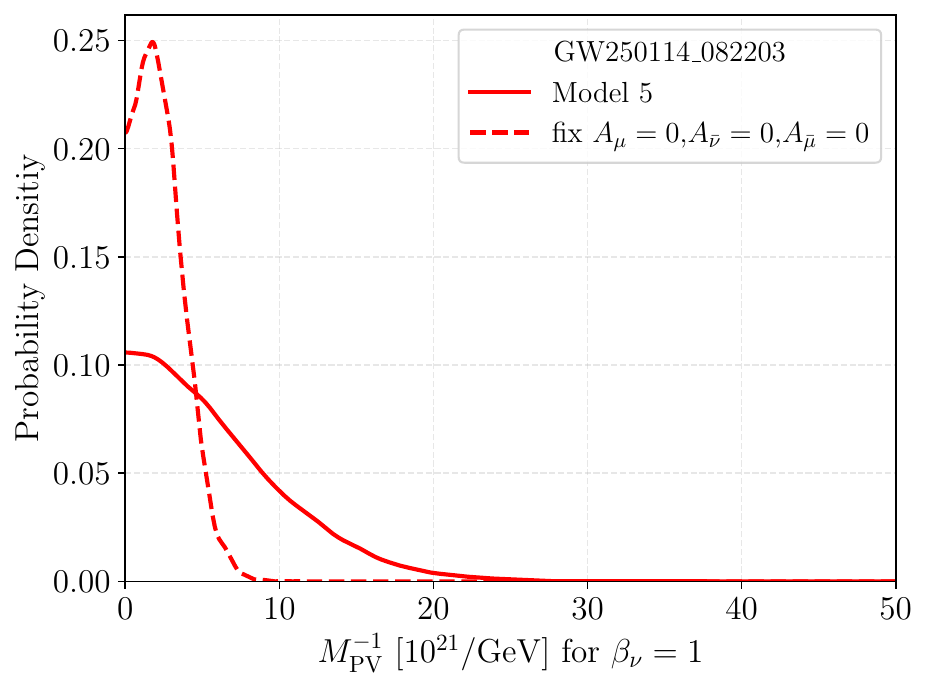}
  \includegraphics[width=0.235\textwidth]{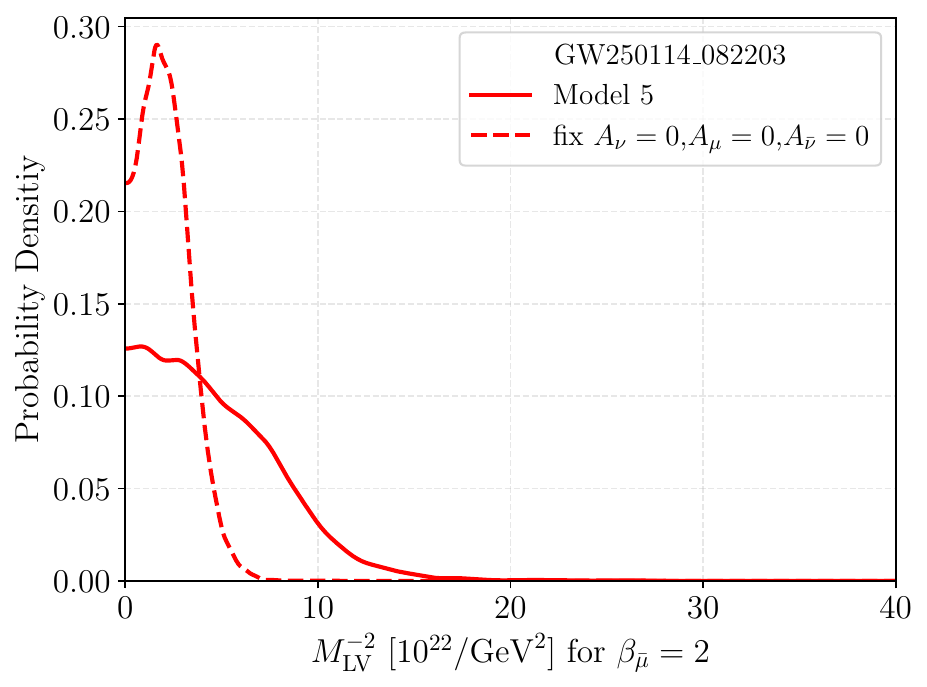}
  \includegraphics[width=0.235\textwidth]{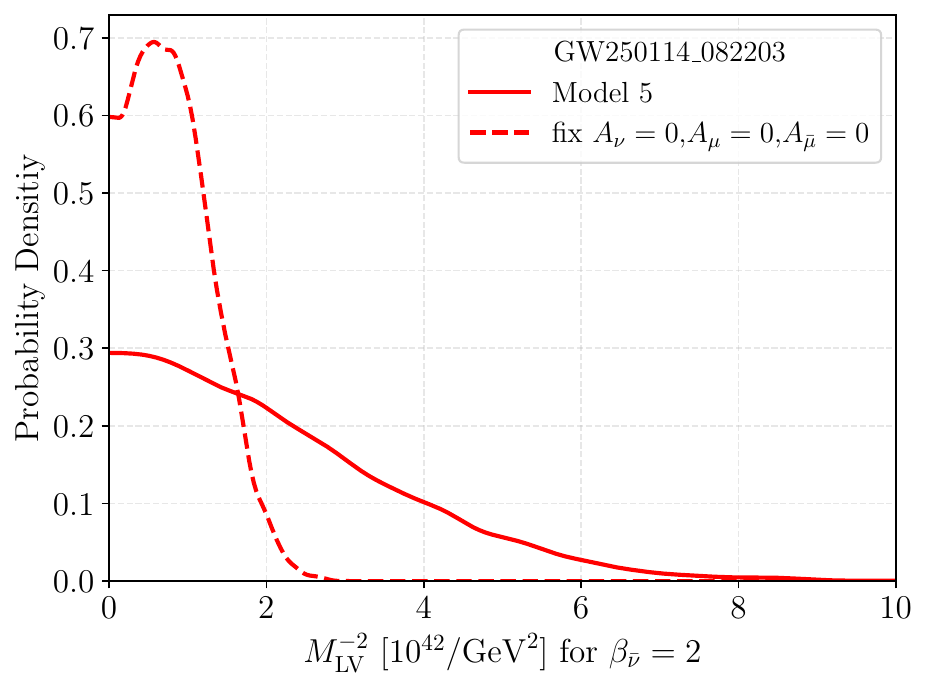}
  \caption{The same as Figure.\ref{fig:fig}, but for $M_{\mathrm{PV}}^{-\beta_\nu}, M_{\mathrm{PV}}^{-\beta_\mu}$, $M_{\mathrm{LV}}^{-\beta_{\bar{\nu}}}$, and $M_{\mathrm{LV}}^{-\beta_{\bar{\mu}}}$ constrained from the analyses of GW250114\_082203.}
  \label{fig:B1}
\end{figure}

\begin{figure*}[t]
  \centering
  \includegraphics[width=0.6\linewidth]{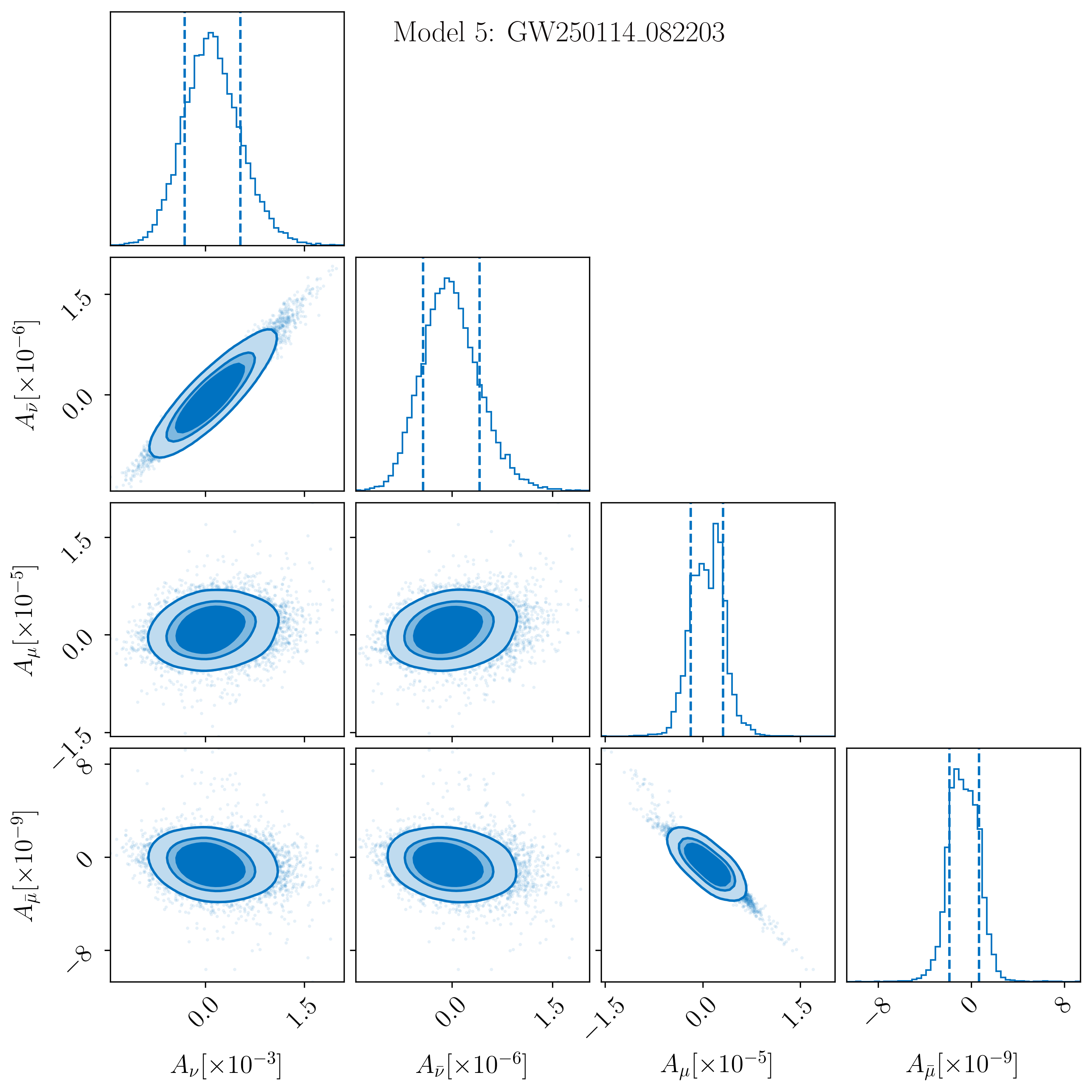}
  \caption{The same as Figure.\ref{fig:2Dcorner}, but for the constraints on ($A_{\nu}$, $A_{\mu}$, $A_{\bar{\nu}}$, $A_{\bar{\mu}}$) derived from the four-parameter analysis of GW250114\_082203.}
  \label{fig:B2}
\end{figure*}

{Some of the inference results for Model 5 are shown in Figure.\ref{fig:B1} and Figure.\ref{fig:B2}. One can see from Figure.\ref{fig:B1} that the multi-parameter analysis obtains weaker constraints than that of single-parameter analyses, which is similar to the outcomes of Model 4. In addition, Figure.\ref{fig:B2} shows that $A_{\nu}$ correlates with $A_{\bar{\nu}}$, and $A_{\mu}$ correlates with $A_{\bar{\mu}}$ in the inference. The reason is that although $A_{\nu}$ and $A_{\mu}$ introduce birefringences of GWs, the viewing angle of GW250114\_082203 is $\sim 0.78$, which makes the left-handed polarization modes much stronger than the right-handed modes. As a consequence, $A_{\bar{\nu}}$ ($A_{\bar{\mu}}$) degenerates with $A_{\nu}$ ($A_{\mu}$) because the detected GW signal is dominated by just one mode. The converted constraints on $M_{\rm PV}$ and $M_{\rm LV}$ are also listed in Table~\ref{tab:tab2}.}

%\clearpage

\end{document}